\documentclass[10pt,letterpaper,twocolumn,ruled,vlined,linesnumbered,twocolumn,twoside]{IEEEtran}
\usepackage[utf8]{inputenc}
\usepackage{array}
\usepackage{verbatim}
\usepackage{multirow}
\usepackage{algorithm2e}
\usepackage{amsmath}
\usepackage{amsthm}
\usepackage{amssymb}
\usepackage{graphicx}

\makeatletter

\pdfpageheight\paperheight
\pdfpagewidth\paperwidth

\providecommand{\tabularnewline}{\\}

\theoremstyle{plain}
\newtheorem{thm}{\protect\theoremname}
\theoremstyle{plain}
\newtheorem{prop}[thm]{\protect\propositionname}

\usepackage{amsmath}
\usepackage{amsfonts}
\usepackage{amssymb}
\usepackage{amsthm}

\usepackage{graphicx}

\usepackage{multirow}
\usepackage{array}
\usepackage{slashbox}
\usepackage{makecell}
\usepackage{threeparttable}

\usepackage{rotating}

\usepackage{pict2e}
\usepackage{cases}
\usepackage{enumitem}
\usepackage{footnote}
\makesavenoteenv{tabular}
\makesavenoteenv{table}

\usepackage{floatrow}
\usepackage{color}
\usepackage{cite} 

\usepackage{url}
\usepackage{adjustbox}

\newcommand{\figref}{Fig. }
\newcommand{\tabref}{Table }
\newcommand{\secref}{Section }
\newcommand{\algref}{Algorithm }
\newcommand{\prpref}{Proposition }

\title{High Throughput Polar Decoding Using Two-Staged Adaptive Successive Cancellation List Decoding}
\author{
ChenYang~Xia,~\IEEEmembership{Student Member,~IEEE,}
YouZhe~Fan,~\IEEEmembership{Member,~IEEE,}
Chi-ying~Tsui,~\IEEEmembership{Senior Member,~IEEE}
\thanks{C.-Y. Xia, and C.-Y. Tsui 
are with the Department of Electronic and Computer Engineering, 
Hong Kong University of Science and Technology, Kowloon, Hong Kong 
(e-mail: cxia@connect.ust.hk; eetsui@ust.hk).}
\thanks{Y.-Z. Fan is with MaxLinear, Carlsbad, CA, USA
(e-mail: jasonfan@connect.ust.hk).}
\thanks{This paper will be presented in part at the IEEE International Conference on Circuits and Systems, Sapporo, Japan, May 2019.}
}
\@ifundefined{showcaptionsetup}{}{%
 \PassOptionsToPackage{caption=false}{subfig}}
\usepackage{subfig}
\makeatother

\providecommand{\propositionname}{Proposition}
\providecommand{\theoremname}{Theorem}

\begin{document}
\maketitle


\bstctlcite{IEEEexample:BSTcontrol_jrl}

\begin{abstract}
Polar codes are the first class of capacity-achieving forward error
correction (FEC) codes. They have been selected as one of the coding
schemes for the 5G communication systems due to their excellent error
correction performance when successive cancellation list (SCL) decoding
with cyclic redundancy check (CRC) is used. A large list size is necessary
for SCL decoding to achieve a low error rate. However, it impedes
SCL decoding from achieving a high throughput as the computational
complexity is very high when a large list size is used. In this paper,
we propose a two-staged adaptive SCL (TA-SCL) decoding scheme and
the corresponding hardware architecture to accelerate SCL decoding
with a large list size. Constant system latency and data rate are
supported by TA-SCL decoding. To analyse the decoding performance
of TA-SCL, an accurate mathematical model based on Markov Chain is
derived, which can be used to determine the parameters for practical
designs. A VLSI architecture implementing TA-SCL decoding is then
proposed. The proposed architecture is implemented using UMC 90nm
technology. Experimental results show that TA-SCL can achieve throughputs
of 3.00 and 2.35 Gbps when the list sizes are 8 and 32, respectively,
which are nearly 3 times as that of the state-of-the-art SCL decoding
architectures, with negligible performance degradation on a wide signal-to-noise
ratio (SNR) range and small hardware overhead.
\end{abstract}


\section{Introduction\label{sec:introduction}}

\IEEEPARstart{S}{ince} they were invented by Ar\i can in 2009, polar
codes \cite{earikan_bilkent_tit_2009_polar} have attracted much research
interest due to their excellent error correction performance. Polar
codes decoded by successive cancellation (SC) decoding provably achieve
channel capacity for symmetric binary-input, discrete, memoryless
channels (B-DMC) when their code lengths are approaching infinity
\cite{ital_ucsd_tit_2013_construction}. However, as the source word
is recovered bit by bit in the SC decoding process, the decoding latency
for long polar code is large \cite{cleroux_mcgill_tsp_2013_semiparallel,yzfan_hkust_tsp_2014_effps}.
Numerous fast SC architectures have been proposed to improve the decoding
latency \cite{gsarkis_mcgill_jsac_2014_fast,czhang_umn_icc_2012_lookahead}.
On the other hand, the error correction performance of SC decoding
is not satisfactory when it is used for practical polar codes with
short to medium code lengths \cite{ital_ucsd_tit_2015_list}, such
as the channel codes for 5G communication systems \cite{3gpp_3gpp_ran087_2016_5g}.
Thus, successive cancellation list (SCL) decoding \cite{kchen_bupt_iet_2012_lscd,ital_ucsd_tit_2015_list}
has been proposed to improve the error correction performance of polar
codes. However, it has a large computational complexity and latency
overhead.

In SCL decoding, $\mathcal{L}$ concurrently-executed SC decodings
are used to keep $\mathcal{L}$ candidates of decoded vectors, where
$\mathcal{L}$ is called the list size. Compared with SC decoding,
SCL decoding has better error correction performance as the source
word is possible to be kept in the list even when a decoding error
happens. Moreover, by concatenating polar codes with cyclic redundancy
check (CRC) codes \cite{bli_huawei_cl_2012_crc,kniu_bupt_cl_2012_crc},
the valid output vector is selected according to the CRC checksums
after decoding. Consequently, SCL decoding significantly out-performs
SC decoding for polar codes in error correction performance. Polar
codes using SCL decoding with $\mathcal{L}\geq16$ even out-perform
low-density parity-check (LDPC) \cite{rggallager_mit_book_1963_ldpc}
and turbo \cite{cberrou_bretagne_icc_1993_turbo} codes using iterative
decoding \cite{kniu_bupt_icc_2013_beyondturbo,yzfan_hkust_icassp_2015_dts},
and hence short polar codes have been elected as one of the coding
schemes in the coming 5G enhanced mobile broadband (eMBB) standard
\cite{3gpp_3gpp_ran087_2016_5g}.

\begin{figure}
\includegraphics[width=8.6cm]{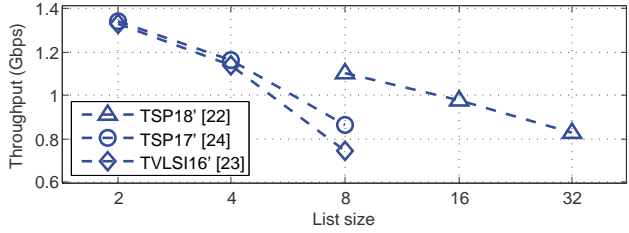}

\caption{Throughputs versus list sizes of various VLSI architectures of SCL
decoders synthesized with or scaled to 90nm CMOS technology.}

\label{fig:soa_thp}
\end{figure}

Aiming at increasing the decoding throughput of polar codes, VLSI
architectures of SCL decoding becomes a popular research topic \cite{abalatsoukas_epfl_tsp_2015_llrlscd,pgiard_epfl_jetcas_2017_polarbear,byuan_umn_tvlsi_2015_sclmbd,crxiong_lehigh_tvlsi_2016_sclmm,crxiong_lehigh_tsp_2016_symbol,yzfan_hkust_jsac_2016_sedts,cxia_hkust_tsp_2018_largelist,jlin_lehigh_tvlsi_2016_highthpt,sahashemi_mcgill_tsp_2017_fastflexible,sahashemi_mcgill_tcasi_2016_ssclspc,gsarkis_mcgill_jsac_2016_sclfast,jlin_lehigh_tvlsi_2015_efficient}.
Compared with single SC decoding, SCL decoding has latency overhead
because of the need of executing list management (LM) \cite{abalatsoukas_epfl_tsp_2015_llrlscd}.
During the decoding process of a bit, the $\mathcal{L}$ survival
paths will be expanded to $2\mathcal{L}$ paths as all of them are
possible candidates of the partial decoded vectors. LM is executed
to select the $\mathcal{L}$ best paths to keep. Basically, LM needs
to solve a radix-$2\mathcal{L}$ sorting problem which has a computational
complexity of $\mathcal{O}(\mathcal{L}^{2})$ \cite{abalatsoukas_epfl_iscas_2015_sorting}.
To minimize the latency overhead brought by LM, the most popular optimization
schemes used in state-of-the-art hardware architectures are decoding
multi-bit sub-codes at the same time so that fewer LM operations are
needed. A sub-code can be either fixed-length \cite{cxia_hkust_tsp_2018_largelist,byuan_umn_tvlsi_2015_sclmbd,crxiong_lehigh_tvlsi_2016_sclmm,crxiong_lehigh_tsp_2016_symbol,yzfan_hkust_jsac_2016_sedts}
or matching a special code pattern with variable length \cite{jlin_lehigh_tvlsi_2015_efficient,jlin_lehigh_tvlsi_2016_highthpt,gsarkis_mcgill_jsac_2016_sclfast,sahashemi_mcgill_tcasi_2016_ssclspc,sahashemi_mcgill_tsp_2017_fastflexible}.
Besides, the sorting algorithm itself can be simplified \cite{abalatsoukas_epfl_iscas_2015_sorting,bykong_kaist_tcasii_2016_sorting,vbioglio_huawei_cl_2017_sorting}.
An approximate sorting algorithm called double thresholding scheme
(DTS) was proposed in our previous work \cite{yzfan_hkust_icassp_2015_dts,yzfan_hkust_jsac_2016_sedts,cxia_hkust_tsp_2018_largelist}.
It simplifies the sorting complexity to $\text{\ensuremath{\mathcal{O}}}(\mathcal{L})$
with the help of two run-time generated thresholds. The corresponding
VLSI architecture supports a list size up to 32 \cite{cxia_hkust_tsp_2018_largelist}
so as to achieve an excellent decoding performance. However, as shown
in \figref \ref{fig:soa_thp}, state-of-the-art SCL decoding architectures
suffer from a severe throughput degradation when the list size is
increased. It is because a larger list size causes larger computational
complexity for LM and hence the critical path delay of the SCL decoding
architectures increases. 

In the iterative decoding of LDPC codes, the number of iterations
for each frame to converge is not fixed so that the decoding speed
can be increased by adaptively assigning different decoding iterations
for different frames \cite{gbosco_polito_cl_2005_buffer,mrovini_unipi_vtc_2007_buffer,slsweatlock_caltech_itaw_2008_buffer}.
Similarly, to increase the decoding speed of polar codes so that they
can be comparable with those of LDPC and turbo codes, adaptive SCL
(A-SCL) decoding was proposed in \cite{bli_huawei_cl_2012_crc}, in
which the list size is adaptive. Specifically, a codeword is first
decoded by a single SC decoding. If the decoded vector cannot pass
CRC, the codeword will be decoded by SCL with a doubled list size.
This process is iterated until a valid output is obtained or a predefined
maximal list size $\mathcal{L}_{\text{max}}$ is reached. Experimental
results in \cite{bli_huawei_cl_2012_crc} show that A-SCL with $\mathcal{L}_{\text{max}}$
has an equivalent error correction performance as that of SCL decoding
with $\mathcal{L}$=$\mathcal{L}_{\text{max}}$, and the average list
size $\bar{\mathcal{L}}\ll\mathcal{L}_{\text{max}}$ as most of the
codewords can be decoded by SCL with $\mathcal{L}\ll\mathcal{L}_{\text{max}}$.
According to the relationship between list size and throughput as
shown in \figref \ref{fig:soa_thp}, the reduction on $\bar{\mathcal{L}}$
increases the average throughput of hardware polar decoder. Nevertheless,
the decoding latency of each codeword in A-SCL is different. This
is not an issue for a software decoder such as the one in \cite{gsarkis_mcgill_jsac_2016_sclfast}.
However, a directly-mapped A-SCL hardware architecture cannot support
applications that need a constant system latency and transmission
data rate, such as the digital baseband in a communication system.

In this work, we will introduce how to accelerate polar decoding on
hardware with the help of A-SCL decoding. The paper is an extension
of our previous work \cite{cxia_hkust_iscas_2019_adaptive}. Here,
the main contributions of this work are summarized as follows:
\begin{itemize}
\item A hardware-friendly two-staged adaptive SCL (TA-SCL) algorithm is
proposed, which can achieve a high throughput with constant transmission
data rate and system latency. To analyse the error correction performance
of TA-SCL, a mathematical model on B-DMC is developed based on Markov
chain. The model is an extension from our previous work \cite{cxia_hkust_iscas_2019_adaptive}
such that the speed gain achieved by TA-SCL is not restricted to an
integer multiple but can be any rational multiple.
\item The relationships between the error correction performance and design
parameters are studied, and a method of how to select the design parameters
is introduced.
\item A hardware architecture of TA-SCL is developed based on the proposed
model. The memory usage is analysed and the corresponding timing schedule
is presented. A low-latency SCL (LL-SCL) architecture combining several
existing low-latency decoding schemes is introduced to satisfy the
high requirement of a component SCL decoder in the TA-SCL decoder
architecture. 
\item Experimental results show that the throughput is about three times
that of state-of-the-art SCL decoding architecture \cite{cxia_hkust_tsp_2018_largelist}
with negligible performance degradation and small hardware overhead.
\end{itemize}
The rest of this paper is organized as follows. In \secref \ref{sec:preliminaries},
the background knowledge of polar codes and the decoding algorithms
will be reviewed. In \secref \ref{sec:ta-scl}, the algorithm of
TA-SCL and its analytical model will be introduced. The relationships
between the error correction performance and design parameters will
also be analysed. In \secref \ref{sec:hw_arc}, the hardware architecture
of the TA-SCL decoder will be introduced. Finally, simulation and
implementation results of the hardware-based TA-SCL will be presented
in \secref \ref{sec:experiment}, and conclusions will be given in
\secref \ref{sec:conclusion}.%

\section{Preliminaries\label{sec:preliminaries}}

\subsection{Polar Codes\label{subsec:polar-codes}}

Polar codes \cite{earikan_bilkent_tit_2009_polar} are a kind of linear
block codes of length $N$. Without loss of generality, we assume
$N$=$2^{n}$ in this work, where $n$ is an integer. Let $\textbf{u}_{N}$
and $\textbf{x}_{N}$ be the input source word and the output codeword
of an $N$-bit binary frame, respectively, and the encoding process
can be simply expressed by $\textbf{x}_{N}=\textbf{u}_{N}\cdot\text{\textbf{F}}^{\otimes n},$
in which $\text{\textbf{F}}^{\otimes n}$ is called the generator
matrix that equals to the $n^{th}$ Kronecker power of the polarization
matrix $\textbf{F}$=$\left[\begin{array}{cc}
1 & 0\\
1 & 1
\end{array}\right]$. Due to the polarization effect, each bit in $\textbf{u}_{N}$ has
a different reliability. An information set $\mathcal{A}$ is determined
by finding the $K$ most reliable bits. These $K$ bits, called information
bits, are used to transmit information. The complement of $\mathcal{A}$,
$\mathcal{A}^{c}$, is defined as the frozen set, in which the bits
are called the frozen bits and set to 0. If an $r$-bit CRC code is
used, the last $r$ information bits are used to transmit the checksum
generated from the other $K$-$r$ information bits, and the code
rate of polar codes is $R=\frac{K-r}{N}$.

\subsection{Successive Cancellation Decoding\label{subsec:sc}}

\begin{figure}
\includegraphics[width=8.6cm]{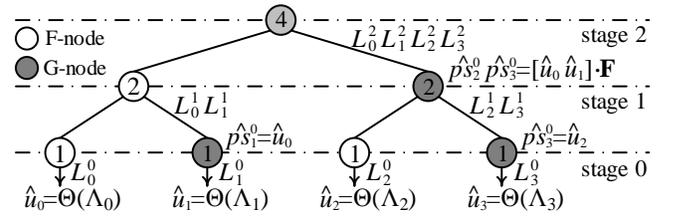}

\caption{Scheduling tree of SC decoding for polar codes with $N=4$.}
\label{fig:schedule_tree}
\end{figure}
Successive cancellation decoding is a basic decoding algorithm of
polar codes and has a low computational complexity of $\mathcal{O}(N\log N)$.
Its decoding process is usually represented by a scheduling tree.
An example for an $N$=4 polar code is shown in \figref \ref{fig:schedule_tree}.
It is a full binary tree with $n$+1 stages. The operands of SC decoding
are the log-likelihood ratios (LLRs). The $i^{th}$ LLR at stage $s$
is denoted as $L_{i}^{s}$, where $i\in[0,N-1]$ and $s\in[0,n]$.
Specifically, $L_{i}^{n}$s are the channel LLRs that are inputted
to the tree from the root node at stage $n$. $L_{i}^{0}$s are LLRs
corresponding to the $N$ leaf nodes and the hard decision of $u_{i}$,
denoted as $\hat{u_{i}}$, is made according to
\begin{equation}
\hat{u}_{i}=\Theta(\Lambda_{i})=\begin{cases}
0, & \textrm{if}\,\Lambda_{i}>0\text{ or }i\in\mathcal{A}^{c},\\
1, & \text{otherwise.}
\end{cases}\label{eq:sgn_func}
\end{equation}
where $\Lambda_{i}$=$L_{i}^{0}$. To obtain these $\Lambda_{i}$s,
the nodes in the scheduling tree are calculated as follows. A pair
of sibling nodes at stage $s$ share $2^{s+1}$ LLR inputs from stage
$s$+1 and both of them execute $2^{s}$ calculations in parallel.
The left and right sibling nodes (denoted as F- and G-nodes) calculate
the following F- and G-functions, respectively:
\begin{align}
L_{\text{F}}(L_{a},L_{b}) & =(-1)^{\Theta(L_{a})\oplus\Theta(L_{b})}\cdot\text{min}(|L_{a}|,|L_{b}|),\label{eq:f_func}\\
L_{\text{G}}(L_{a},L_{b},\hat{ps}) & =(-1)^{\hat{ps}}L_{a}+L_{b},\label{eq:g_func}
\end{align}
where $L_{a}$ and $L_{b}$ are the two input LLRs%
{} and $\hat{ps}$ for the G-function is a binary bit called partial-sum.
\eqref{eq:f_func} is a hardware-friendly version of F-function proposed
in \cite{cleroux_mcgill_tsp_2013_semiparallel}. For a G-node at stage
$s$, its partial-sums are obtained by
\begin{equation}
[\hat{ps}_{j+1}^{s},..,\hat{ps}_{j+2^{s}}^{s}]=[\hat{u}_{j-2^{s}+1},..,\hat{u}_{j}]\cdot\textbf{F}^{\otimes s},\label{eq:ps}
\end{equation}
where $\hat{u}_{j}$ is the last decoded bit. According to \eqref{eq:ps},
the partial-sums of a G-node has data dependancy on the $2^{s}$ decoded
bits rooted at its sibling F-node. Thus, it can be seen that the decoding
process of SC decoding follows a depth-first traversal of the scheduling
tree.

\subsection{Successive Cancellation List Decoding\label{subsec:scl}}

\begin{figure}
\subfloat[Traditional SCL decoding.]{\includegraphics[scale=0.98]{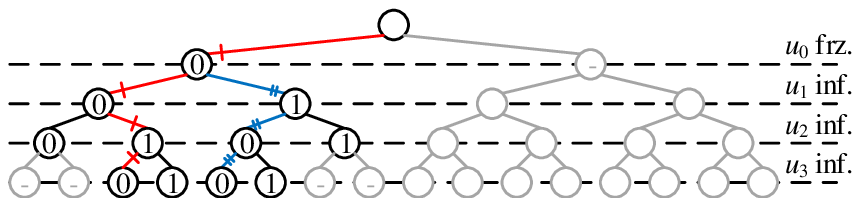}

\label{fig:decoding_tree}}

\subfloat[SCL decoding with MBD.]{\includegraphics[scale=0.98]{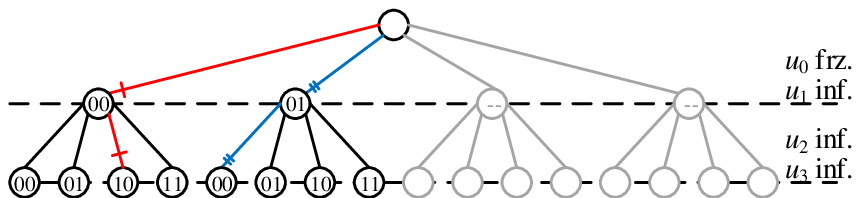}

\label{fig:decoding_tree_mbd}}

\subfloat[SCL decoding with SND.]{\includegraphics[scale=0.98]{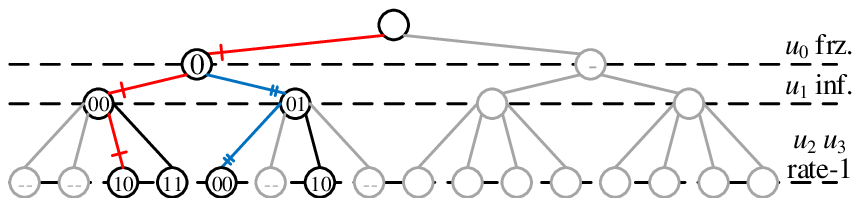}

\label{fig:decoding_tree_snd}}\caption{Decoding tree of different SCL decodings for a polar code with $N=4$
and $\mathcal{L}=2$.}
\end{figure}
SCL decoding was proposed in \cite{ital_ucsd_tit_2015_list,kchen_bupt_iet_2012_lscd}.
It has a significant performance gain over single SC decoding. Its
decoding process can be regarded as a search problem on a binary decoding
tree of depth-$N.$ \figref \ref{fig:decoding_tree} shows a decoding
tree for a polar code of $N$=4. The $i^{th}$ source bit $u_{i}$,
which corresponds to the $i^{th}$ leaf node in the scheduling tree,
is mapped to the nodes at depth $i$+1 in the decoding tree. A path
from the root node to a leaf node represents a candidate of decoded
vector. For a parent node at depth $i$, its left and right children
correspond to two different expansions of the partial decoded path
with $u_{i}$=0 and 1, respectively. For example, the paths marked
with single and double crosslines in \figref \ref{fig:decoding_tree}
represent decoding vectors 0010 and 0100, respectively.

The decoding process begins from the root node. When the decoding
process reaches a frozen bit $u_{i}$, such as $u_{0}$ in \figref
\ref{fig:decoding_tree}, the sub-tree rooted at the right child is
pruned (marked with light colour) as it does not contain any valid
path, and the number of valid paths is unchanged. Otherwise, if $u_{i}$
is an information bit, the valid decoded paths are expanded to both
sibling nodes and the number of valid paths in the list doubles. The
number of the path candidates increases exponentially with respect
to the number of decoded information bits. When a practical code length
is used, the computational complexity will be too high to be implemented
after a few bits are decoded. To limit the computational complexity,
an LM operation is executed at each new depth to keep the number of
survival paths to a predefined value $\mathcal{L}$ which is called
the list size. In \figref \ref{fig:decoding_tree}, the lines with
dark colours represent the paths that have been expanded during LM
and those with crosslines represent the paths kept after LM.

The criterion of selecting survival paths during LM is their reliability
measured by path metrics (PM). We denote the path metric of a path
$l$ ($l\in[0,\mathcal{L}-1]$) after the decoding of bit $u_{i}$
as $\gamma_{i+1}^{k}$ where $k\in\{2l,2l$+1\}. The PM is initialized
as $\gamma_{0}^{0}$=0 and is updated based on bit-wise accumulation
as \cite{abalatsoukas_epfl_tsp_2015_llrlscd}
\begin{equation}
\begin{cases}
\gamma_{i+1}^{2l}=\gamma_{i}^{l}, & \textrm{where}~\hat{u}_{i}^{2l}=\Theta\left(\Lambda_{i}^{l}\right),\\
\gamma_{i+1}^{2l+1}=\gamma_{i}^{l}+|\Lambda_{i}^{l}|, & \textrm{where}~\hat{u}_{i}^{2l+1}=1-\Theta\left(\Lambda_{i}^{l}\right).
\end{cases}\label{eq:pmu_aprx}
\end{equation}
Similar as \eqref{eq:f_func}, \eqref{eq:pmu_aprx} is a hardware-friendly
version of PM update. For a frozen bit, only one of \eqref{eq:pmu_aprx}
that satisfies $\hat{u_{i}}$=0 will be computed and the number of
paths remains to be $\mathcal{L}$. As mentioned above, only the $\mathcal{L}$
left children nodes will be kept. Otherwise, both equations in \eqref{eq:pmu_aprx}
will be computed and the number of paths is doubled. After that, a
list pruning operation will be executed, where all the $2\mathcal{L}$
PMs are sorted and the $\mathcal{L}$ paths with the smaller PMs will
be kept in the list.

Recently, a variety of algorithms have been proposed to reduce the
decoding latency of SCL by decoding multiple bits and executing their
LM operation for only once. The algorithms in the literature can be
divided into two different classes, multi-bit decoding (MBD) \cite{cxia_hkust_tsp_2018_largelist,byuan_umn_tvlsi_2015_sclmbd,crxiong_lehigh_tvlsi_2016_sclmm,crxiong_lehigh_tsp_2016_symbol}
and special node decoding (SND) \cite{jlin_lehigh_tvlsi_2015_efficient,jlin_lehigh_tvlsi_2016_highthpt,gsarkis_mcgill_jsac_2016_sclfast,sahashemi_mcgill_tcasi_2016_ssclspc,sahashemi_mcgill_tsp_2017_fastflexible}.
MBD decodes $M=2^{m}$ bits simultaneously, where $M$ is a fixed
and predefined value. The decoding tree of MBD is modified to a full
$2^{M}$-ary tree as shown in \figref \ref{fig:decoding_tree_mbd},
in which $M$=2. LM is still executed at each depth of this tree for
each $M$ bits. SND, on the other hand, runs simplified LM algorithms
for variable-length sub-codes that matches special code patterns.
\figref \ref{fig:decoding_tree_snd} shows the decoding tree modified
for SND, in which rate-1 sub-code, a sub-code with only information
bits, is used to simplify the decoding. Fewer paths are expanded from
each survival path in SND and hence the computational complexity is
reduced.%

\subsection{Adaptive SCL Decoding for Polar Codes\label{subsec:a-scl}}

{\small{}}
\begin{algorithm}[t]
{\small{}\caption{Adaptive SCL with CRC.}
\label{alg:a-scl}}{\small \par}

{\small{}\textbf{Input:} $N$ channel LLRs; \textbf{Initial:} $\mathcal{L}=1$\;
\While{\textnormal{1}}{
	SCL decoding with $\mathcal{L}$: codeword from channel\;
	\If{more than one paths pass CRC \textnormal{or} $\mathcal{L}==\mathcal{L}_{\text{max}}$}{Output the most reliable path; Break;}
	\Else{$\mathcal{L}=2\cdot\mathcal{L}$; // $\mathcal{L}=\mathcal{L}_{\text{max}}$ for simplified A-SCL}
}	}{\small \par}
\end{algorithm}
\begin{table}[t]
\caption{Average list size of the $(N,K,r)$=$(1024,512,24)$ polar code decoded
by A-SCL with $\mathcal{L}_{\text{max}}$=32}
\label{tab:average_list_size_steps}

{\small{}}%
\begin{tabular}{c|c|c|c|c|c}
\hline 
\multicolumn{1}{c|}{{\small{}SNR}} & {\small{}1.2} & {\small{}1.4} & {\small{}1.6} & {\small{}1.8} & {\small{}2.0}\tabularnewline
\hline 
{\small{}Original \cite{bli_huawei_cl_2012_crc}} & {\small{}4.11} & {\small{}2.49} & {\small{}1.67} & {\small{}1.30} & {\small{}1.13}\tabularnewline
\hline 
{\small{}Simplified \cite{gsarkis_mcgill_jsac_2016_sclfast}} & {\small{}18.55} & {\small{}13.52} & {\small{}9.06} & {\small{}5.74} & {\small{}3.53}\tabularnewline
\hline 
\end{tabular}{\small \par}
\end{table}

Adaptive SCL with CRC was proposed in \cite{bli_huawei_cl_2012_crc}
and the algorithm is summarised in \algref \ref{alg:a-scl}. Each
time, a new codeword which contains $N$ LLRs is inputted for decoding.
A-SCL starts with an SCL of $\mathcal{L}$=1, i.e., a single SC. If
there are some decoded vectors that pass CRC at the end of decoding,
the one with the highest reliability is chosen as output. Otherwise,
the list size is doubled and the codeword is decoded again by an SCL
with the new list size. Usually, a predefined $\mathcal{L}_{\text{max}}$
is used to limit the computational complexity, that is, after the
decoding using an SCL with $\mathcal{L}_{\text{max}}$, the decoding
terminates even when there is no valid candidate. From \cite{bli_huawei_cl_2012_crc},
the error correction performance of A-SCL is the same as that of an
SCL with $\mathcal{L}_{\text{max}}$. At the same time, as most of
the valid decoded vectors can be obtained using SCL with smaller list
sizes, the average list size $\bar{\mathcal{L}}$ of A-SCL is much
smaller than $\mathcal{L}_{\text{max}}$. An example for a polar code
of $(N,K,r)$=$(1024,512,24)$ polar code is shown in \tabref \ref{tab:average_list_size_steps},
in which $\bar{\mathcal{L}}\ll\mathcal{L}_{\text{max}}$=32 . As SCL
with a smaller list size has a higher decoding speed, the average
decoding speed of A-SCL is much higher than that of SCL.

However, if we directly implement the A-SCL algorithm in hardware,
the following issues need to be addressed:
\begin{itemize}
\item In an A-SCL decoding, different codewords may be decoded by SCL with
different list sizes and SCL with a larger list size has a much larger
latency. The decoding latency varies from frame to frame, so the system
latency is not fixed. Because of that, a directly-mapped architecture
may not be able to support applications that need to have a constant
transmission data rate and latency, such as the channel coding blocks
in communication systems.
\item A directly-mapped architecture is required to support multiple SCL
decodings with list sizes $\mathcal{L}$ ranging from 1 to $\mathcal{L}_{\text{max}}$.
This increases the design effort and also the hardware complexity.
A simplified A-SCL decoding was proposed for accelerating software
polar decoding \cite{gsarkis_mcgill_jsac_2016_sclfast}, in which
only one single SC and one SCL with $\mathcal{L_{\text{max}}}$ is
used. However, as shown in \tabref \ref{tab:average_list_size_steps},
its $\bar{\mathcal{L}}$ is larger than that of the original A-SCL,
which means the achievable throughput gain is much less than that
of the original A-SCL decoding. Moreover, simplified A-SCL does not
support constant transmission data rate as well.
\item The throughput of A-SCL is not a constant under different channel
conditions. As shown in \tabref \ref{tab:average_list_size_steps},
the average list sizes of A-SCL decoding increase when the channel
condition deteriorates as more frames need to be decoded multiple
times, which indicates the throughput could be affected accordingly.
It is also shown in \cite{gsarkis_mcgill_jsac_2016_sclfast} that
a software-based simplified A-SCL decoder suffers from a 20x throughput
reduction when the signal-to-noise ratio (SNR) is reduced by 1 dB.
To adjust the data rate according to the channel condition, the transmitter
side in a real system needs to know the channel SNR in real time,
which increases the difficulty of system implementation.
\end{itemize}
To design a hardware-friendly A-SCL algorithm, we take reference from
variable-iteration decoders for LDPC codes \cite{gbosco_polito_cl_2005_buffer,mrovini_unipi_vtc_2007_buffer,slsweatlock_caltech_itaw_2008_buffer}.
An additional buffer is employed at the input of the iterative decoder,
in which the newly received codeword can be stored temporarily before
the current decoding is finished. By doing so, the iterative decoder
can use different decoding iterations to achieve decoding convergence
and support constant transmission data rate at the same time. In the
next section, we will propose a two-staged A-SCL decoding which solves
the problems mentioned above with the help of some buffers and is
hence suitable for hardware implementation. %
{} 

\section{Two-staged Adaptive Successive Cancellation Decoding\label{sec:ta-scl}}

\subsection{Algorithm of TA-SCL Decoding\label{subsec:alg_tascl}}

\begin{figure}
\includegraphics[width=8.8cm]{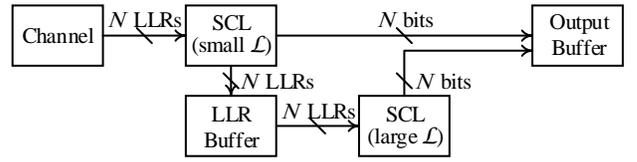}\caption{Block diagram of TA-SCL.}
\label{fig:block_diagram}
\end{figure}
\begin{figure}
\subfloat[The buffer size is infinite. Buffer overflow never happens.]{\includegraphics[width=8.8cm]{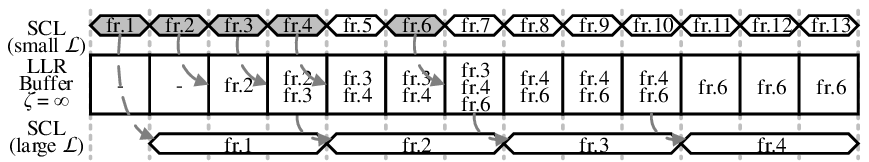}\label{fig:schedule_infinite}}

\subfloat[The buffer size is one (finite). Buffer overflow may happen.]{\includegraphics[width=8.8cm]{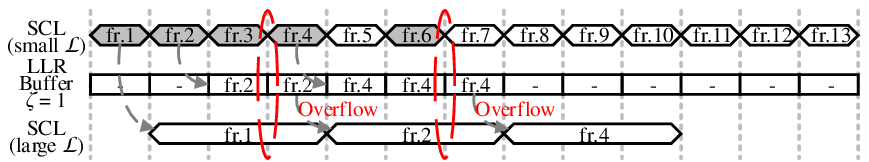}

\label{fig:schedule_finite}}

\caption{Timing schedule of TA-SCL. The codewords in gray cannot be decoded
correctly by $\mathbf{D}_{s}$. The arrows represent data flows.}
\label{fig:schedule}
\end{figure}
As mentioned in \secref \ref{subsec:a-scl}, the average list size
and the error correction performance of A-SCL algorithm follows that
of single SC and SCL with $\mathcal{L}_{\text{max}}$, respectively.
Based on this observation, we propose a hardware-friendly TA-SCL whose
algorithm is described as below.%

The block diagram of TA-SCL and its timing schedule are shown in \figref
\ref{fig:block_diagram} and \figref \ref{fig:schedule_infinite},
respectively. Basically, it includes two SCL decodings, which are
an SCL decoding with small list size (not necessarily to be 1), denoted
as $\mathbf{D}_{s}$, and an SCL decoding with large list size $\mathcal{L}_{\text{max}}$,
denoted as $\mathbf{D}_{l}$. Each input codeword from the channel
is first decoded by $\mathbf{D}_{s}$. If none of the candidates in
the list passes CRC after this decoding, e.g. fr.1 in \figref \ref{fig:schedule_infinite},
the current codeword will be decoded again by $\mathbf{D}_{l}$. This
decoding usually takes longer time than decoding using $\mathbf{D}_{s}$.
Nevertheless, $\mathbf{D}_{l}$ runs concurrently with $\mathbf{D}_{s}$
so that $\mathbf{D}_{s}$ starts decoding the next codeword immediately.
Most of the time, $\mathbf{D}_{s}$ can decode the input codewords
correctly and $\mathbf{D}_{l}$ becomes idle when the current decoding
process finishes. However, if the channel is subject to burst errors,
it is possible that a new codeword cannot be correctly decoded by
$\mathbf{D}_{s}$ and the decoding in $\mathbf{D}_{l}$ has not finished
yet. To deal with this, an LLR buffer is needed to store the LLRs
of the codeword from $\mathbf{D}_{s}$ temporarily, such as fr.2,
fr.3, fr.4 and fr.6 shown in \figref \ref{fig:schedule_infinite}.
An output buffer is also needed to re-order the decoded vectors as
the codeword may be decoded out of order. For example, fr.7\textasciitilde{}fr.13
are stored in the output buffer until the decoding of fr.6 finishes.

The major difference between TA-SCL and A-SCL, either the original
one or the simplified one, is that $\mathbf{D}_{s}$ can decode the
next codeword from the channel input immediately instead of waiting
for $\mathbf{D}_{l}$ to finish decoding the current codeword with
the help of LLR buffer. The continuous running of $\mathbf{D}_{s}$
and LLR storage in the buffer permit the data to be transmitted at
a constant data rate which is equal to the decoding throughput of
$\mathbf{D}_{s}$ regardless of the SNR of the channel, while the
decoding performance is guaranteed by $\mathbf{D}_{l}$. Also, TA-SCL
benefits the hardware complexity as only two SCL decoders need to
be implemented on hardware. The issues mentioned in \secref \ref{subsec:a-scl}
are hence solved. %

\subsection{Performance Bound of TA-SCL Decoding\label{subsec:bound_tascl}}

If we have unlimited buffer resources, the decoding performance of
TA-SCL will be the same as that of $\mathbf{D}_{l}$. However, in
actual hardware implementation, the buffer size is limited and buffer
overflow will happen, as shown in \figref \ref{fig:schedule_finite}.
It happens when a new codeword needs to be stored in the LLR buffer
but the buffer is full and decoding in $\mathbf{D}_{l}$ has not finished
yet. To deal with buffer overflow, either the codeword in $\mathbf{D}_{s}$
or $\mathbf{D}_{l}$ would be thrown away and the incorrect decoding
results from $\mathbf{D}_{s}$ will be used as the final output decoded
vector of the corresponding codeword. Thus, the block error rate (BLER)
of $\mathbf{D}_{\text{TA}}$, denoted as $\epsilon_{\mathbf{D}_{\text{TA}}}$,
is bounded by
\begin{equation}
\epsilon_{l}\leq\epsilon_{\mathbf{D}_{\text{TA}}}<\epsilon_{l}+\text{Pr(Overflow)},\label{eq:bler_d2}
\end{equation}
where the upper limit is the sum of the BLER of $\mathbf{D}_{l}$
and the probability of buffer overflow. Obviously, it is important
to prevent buffer overflow in order to reduce performance loss which
is defined as
\begin{equation}
\delta=\frac{\epsilon_{\textbf{D}_{\text{TA}}}-\epsilon_{l}}{\epsilon_{l}}\cdot100\%.\label{eq:delta_loss}
\end{equation}
In summary, the benefits of TA-SCL on hardware comes at the cost of
error correction performance loss. To obtain the best tradeoff among
performance, hardware usage and throughput, an analytical model of
$\mathbf{D}_{\text{TA}}$ will be introduced next to obtain the relationship
between $\text{Pr(Overflow)}$ and the parameters of $\mathbf{D}_{\text{TA}}$
in the next sub-section. Before that, we define the design parameters
of TA-SCL as follows. %
\begin{itemize}
\item $\mathcal{L}_{s}/\mathcal{L}_{l}$: list sizes of $\mathbf{D}_{s}$
/$\mathbf{D}_{l}$ .
\item $\epsilon_{s}/\epsilon_{l}$: BLERs of $\mathbf{D}_{s}$ /$\mathbf{D}_{l}$
.
\item $t_{s}/t_{l}$: decoding time of each codeword using $\mathbf{D}_{s}$
/$\mathbf{D}_{l}$ .
\item $\beta$: speed gain, i.e., $\frac{t_{l}}{t_{s}}$. In this work,
$\beta$ is not limited to integer value and can be any rational number,
which is given by $\beta=\frac{\beta_{n}}{\beta_{d}},$ where $\beta_{n},\beta_{d}\in\mathbb{Z}^{+}$
and $\beta_{n}\bot\beta_{d}$, i.e., $\beta_{n}$ and $\beta_{d}$
are co-prime. 
\item $\zeta$: size of the LLR buffer, which equals to the number of codewords
that can be stored in the buffer. In this work, we assume $\zeta\geq1$
and $\zeta\in\mathbb{Z}^{+}$.
\end{itemize}
We also denote a TA-SCL decoding whose speed gain is $\beta$ and
buffer size is $\zeta$ as $\mathbf{D}_{\text{TA}}(\beta,\zeta)$.
The TA-SCL decoding in the example shown in \figref \ref{fig:schedule_infinite}
and \ref{fig:schedule_finite} hence can be described as $\mathbf{D}_{\text{TA}}(3,\infty)$
and $\mathbf{D}_{\text{TA}}(3,1)$, respectively, and the corresponding
$\mathbf{D}_{l}$ needs $3t_{s}$ to decode a codeword. 

\subsection{Analytical Model of TA-SCL Based on Markov Chain\label{subsec:model_tascl}}

\begin{figure}
\subfloat[$\mathbf{D}_{\text{TA}}(3,1)$.]{\includegraphics[width=8.8cm]{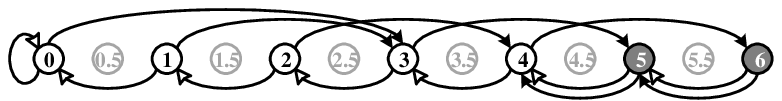}\label{fig:state_transition_3p0}}

\subfloat[ $\mathbf{D}_{\text{TA}}(\frac{5}{2},1)$.]{\includegraphics[width=8.8cm]{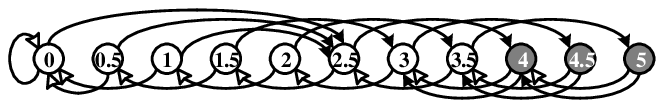}\label{fig:state_transition_2p5}}

\caption{State diagrams of the proposed model. The white and black arrows mean
the frame is decoded correctly and incorrectly, respectively.}

\label{fig:state_transition}
\end{figure}

{} In this sub-section, we model the behavior of $\mathbf{D}_{\text{TA}}(\beta,\zeta)$
on B-DMC. Without loss of generality, we assume the channel is an
additive white Gaussian noise (AWGN) channel. We first introduce the
states that the decoder can operate at. We define the number of codewords
currently stored in the LLR buffer as $i_{\zeta}$ and the remaining
time required to finish the decoding of the current codeword in $\mathbf{D}_{l}$
as $i_{\beta}$ which is an integer multiple of $\frac{t_{s}}{\beta_{d}}$.%
{} Each codeword in the LLR buffer needs $\beta t_{s}$ to decode. Then,
the state of TA-SCL is defined as
\begin{equation}
X_{\tau}=\beta\cdot i_{\zeta}+i_{\beta},\,i_{\zeta}\in[0,\zeta],\,i_{\beta}\in\{\frac{i}{\beta_{d}}|i\in[0,\beta_{n}]\},
\end{equation}
which is actually the total time required to clear the buffer in terms
of $t_{s}$. For a $\mathbf{D}_{\text{TA}}(\beta,\zeta)$, there are
$\mathcal{S}$=$\beta_{n}\zeta$+$\beta_{n}$+1 states in total. The
$\mathcal{S}$ states can be categorized into the following groups
according to whether buffer flow will happen.
\begin{itemize}
\item Hazard states: The states that the LLR buffer is full and the current
codeword decoded by $\mathbf{D}_{l}$ cannot be finished within $t_{s}$,
which means $i_{\zeta}$=$\zeta$ and $i_{\beta}$\textgreater{}$\beta_{d}$.
Buffer overflow will occur if $\mathbf{D}_{s}$ cannot decode the
next codeword correctly. The codeword in $\mathbf{D}_{s}$ will be
thrown away as this allows $\mathbf{D}_{l}$ to decode as many codewords
as possible without any interruption.
\item Safe/Idle states: In contrast with the hazard states, these states
do not have overflow hazard as the LLR buffer has enough space for
a codeword that cannot be correctly decoded by $\mathbf{D}_{s}$.
The transitions in idle states are a little different from those in
safe states in the sense that $\mathbf{D}_{l}$ will finish its decoding
and become idle during this $t_{s}$.
\end{itemize}
\begin{figure}
\adjustbox{valign=b}{\subfloat[An example for $\mathbf{D}_{\text{TA}}(3,1)$.]{\includegraphics[width=4.4cm]{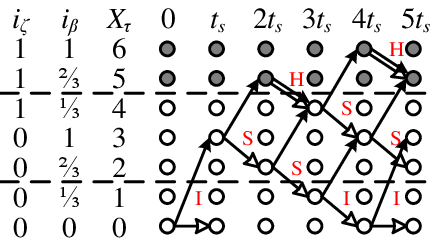}

\label{fig:state_diagram}}}\adjustbox{valign=b}{\subfloat[Summarization of three kinds of states.]{\setlength{\tabcolsep}{1pt}

\begin{tabular}{c|c|>{\centering}p{1.1cm}|>{\centering}p{1.1cm}}
\hline 
 & {\footnotesize{}\# State} & {\footnotesize{}Incorrect decoding} & {\footnotesize{}Correct decoding}\tabularnewline
\hline 
{\footnotesize{}Hazard} & {\footnotesize{}$\beta_{n}$-$\beta_{d}$} & {\footnotesize{}$X_{\tau}$-1} & {\footnotesize{}$X_{\tau}$-1}\tabularnewline
\hline 
{\footnotesize{}Safe} & {\footnotesize{}$\beta_{n}\zeta$} & {\footnotesize{}$X_{\tau}$+$\beta$-1} & {\footnotesize{}$X_{\tau}$-1}\tabularnewline
\hline 
{\footnotesize{}Idle} & {\footnotesize{}$\beta_{d}$+1} & {\footnotesize{}$\beta$} & {\footnotesize{}0}\tabularnewline
\hline 
\end{tabular}\label{fig:state_summary}}

}\caption{States and state transitions of the proposed model. The white and
black arrows mean the frame is decoded correctly and incorrectly,
respectively.}
\end{figure}
We show an example for $\mathbf{D}_{\text{TA}}(3,1)$ in \figref
\ref{fig:state_diagram}, where the black and white arrows represent
the probabilities of $\epsilon_{s}$ and $\epsilon_{s}'$=1-$\epsilon_{s}$,
respectively. The first three columns show $i_{\beta}$, $i_{\zeta}$
and $X_{\tau}$, respectively. Typical transitions from hazard, safe
and idle states are marked with ``H'', ``S'' and ``I'' in the
figure, respectively. The number of these states and their transitions
are summarized in \figref \ref{fig:state_summary}. After defining
these states, we introduce the modeling of TA-SCL by Proposition 1.
\begin{prop}
Decoding with $\mathbf{D}_{\text{TA}}$ is a Markov process and can
be modeled with a Markov chain.
\end{prop}
\begin{IEEEproof}
An input codeword is independent of the codewords inputted at other
time as all the LLRs are random variables with identical and independent
distribution (IID). Thus, the decoding correctness of $\mathbf{D}_{s}$
only depends on the inputs at that time. It actually follows a Bernoulli
distribution which takes the value 1 (error happens) with probability
$\epsilon_{s}$. As all the state transitions depend only on $\epsilon_{s}$,
the next state of $\mathbf{D}_{\text{TA}}(\beta,\zeta)$ only depends
on the current state instead of any earlier state. Hence, Proposition
1 is proved.
\end{IEEEproof}
From Proposition 1, the state diagram of a $\mathbf{D}_{\text{TA}}(\beta,\zeta)$
can be easily obtained by finding out all the possible state transitions
in \figref \ref{fig:state_diagram}. The state diagrams of $\mathbf{D}_{\text{TA}}(3,1)$
and $\mathbf{D}_{\text{TA}}(\frac{5}{2},1)$ are both shown in \figref
\ref{fig:state_transition}. The latter one has some non-integer states
as the remaining time to clear a buffer is an integer multiple of
$\frac{t_{s}}{2}$ instead of $t_{s}$. If $\beta_{n}$ and $\beta_{d}$
are not co-prime, the corresponding model can be simplified. For example,
if $\beta_{n}$=6 and $\beta_{d}$=2 in \figref \ref{fig:state_transition_3p0},
some states (the light gray ones in \figref \ref{fig:state_transition_3p0})
can never be accessed and are redundant in the model. These states
can be removed and the model is the same with the one with $\beta_{n}$=3
and $\beta_{d}$=1. So we only consider the situations that $\beta_{n}\bot\beta_{d}$. 

For further mathematical analysis, we map the state diagram to a transition
matrix $P$ whose size is $\mathcal{S}\times\mathcal{S}$. An element
$P_{x,y}\in P$ $(x,y\in[0,\mathcal{S}$-1{]}) corresponds to the
transition probability from state $x$ to state $y$, i.e.,
\begin{equation}
P_{x.y}=\text{Pr}(X_{\tau+1}=y|X_{\tau}=x),
\end{equation}
where $X_{\tau}$ is the current state and $X_{\tau+1}$ is the next
state. The transition matrix of $\mathbf{D}_{\text{TA}}(3,1)$ mapped
from the state diagram is{\small{}
\begin{equation}
P=\begin{bmatrix}\epsilon_{s}' & 0 & 0 & \epsilon_{s} & 0 & 0 & 0\\
\epsilon_{s}' & 0 & 0 & \epsilon_{s} & 0 & 0 & 0\\
0 & \epsilon_{s}' & 0 & 0 & \epsilon_{s} & 0 & 0\\
0 & 0 & \epsilon_{s}' & 0 & 0 & \epsilon_{s} & 0\\
0 & 0 & 0 & \epsilon_{s}' & 0 & 0 & \epsilon_{s}\\
0 & 0 & 0 & 0 & 1 & 0 & 0\\
0 & 0 & 0 & 0 & 0 & 1 & 0
\end{bmatrix}.\label{eq:trans_mat}
\end{equation}
}The transition matrix $P$ is time-independent according to Proposition
1. To do steady-state analysis for TA-SCL, the following proposition
of the Markov chain model is introduced.
\begin{prop}
The Markov chain of TA-SCL decoding is irreducible (possible to get
to any state from any state) and aperiodic. 
\end{prop}
\begin{IEEEproof}
The proof is given in Appendix A.
\end{IEEEproof}
From Proposition 2, the chain converges to the stationary distribution
regardless of its initial distribution. Suppose that the decoding
begins with $\mathbf{D}_{\text{TA}}$ at state 0, i.e., the initial
distribution $\lambda_{0}$=$[1,0,...,0]$. After $k\cdot t_{s}$
($k\in\mathbb{Z}^{+}$), the probability distribution becomes $\lambda_{k}$=$\lambda_{0}\cdot P^{(k)}$.
Let $P_{\infty}$=$\underset{k\to\infty}{\text{lim}}P^{(k)}$, then
the stationary distribution $\pi$ of $\mathbf{D}_{\text{TA}}$ is%
\begin{equation}
\pi=\lambda\cdot P_{\infty}=[(P_{\infty})_{0,0},...,(P_{\infty})_{0,\beta\zeta+\beta}].\label{eq:ss_dist}
\end{equation}
Actually, all the rows of $P_{\infty}$ are the same, and so the stationary
distribution is irrespective of the initial state $\lambda_{0}$ of
$\mathbf{D}_{\text{TA}}$. Buffer overflow happens when $\mathbf{D}_{\text{TA}}$
is in any hazard state and $\mathbf{D}_{s}$ cannot decode the next
codeword correctly. Thus, the probability of buffer overflow is expressed
as
\begin{eqnarray}
\text{Pr(Overflow)} & = & \epsilon_{s}\cdot\text{Pr(Hazard)}\label{eq:overflow_1}\\
 & = & \epsilon_{s}\cdot\text{Pr}(i_{\zeta}=\zeta\text{ and }i_{\beta}>1)\\
 & = & \epsilon_{s}\cdot\text{Pr}(X_{\tau}>\beta\zeta+1),\\
 & = & \epsilon_{s}\cdot{\displaystyle \sum_{i=\beta\zeta+1+\beta_{n}^{-1}}^{\beta\zeta+\beta}}\pi_{i}.\label{eq:overflow_4}
\end{eqnarray}
This probability of overflow bounds $\epsilon_{\mathbf{D}_{\text{TA}}}$
in \eqref{eq:bler_d2}. It is a function of error correction performance
of $\mathbf{D}_{s}$ $\epsilon_{s}$, speed gain $\beta$ and buffer
size $\zeta$, i.e., $\text{Pr(Overflow)}$=$f(\epsilon_{s},\beta,\zeta)$.
We will use this to analyse the error correction performance of TA-SCL
next.

\subsection{Error Correction Performance of TA-SCL\label{subsec:err_perf_tascl}}

\begin{figure*}
\subfloat[$\mathbf{D}_{\text{TA}}(3,3)$ with $\mathcal{L}_{s}$=1 and 2.]{\includegraphics[width=6cm]{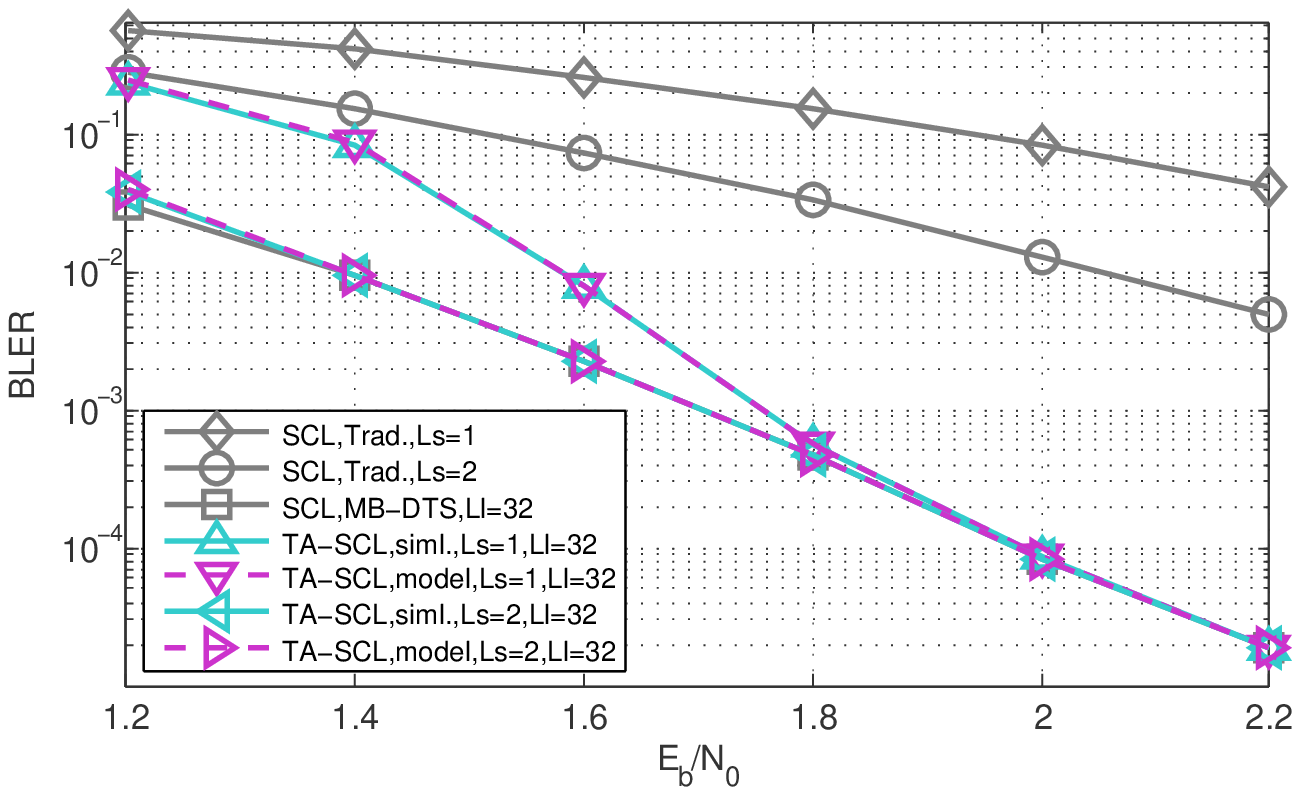}

\label{fig:err_perf_es}}\subfloat[{$\mathbf{D}_{\text{TA}}(\beta,2)$ ($\beta\in[2,4]$) with $\mathcal{L}_{s}$=2.}]{\includegraphics[width=6cm]{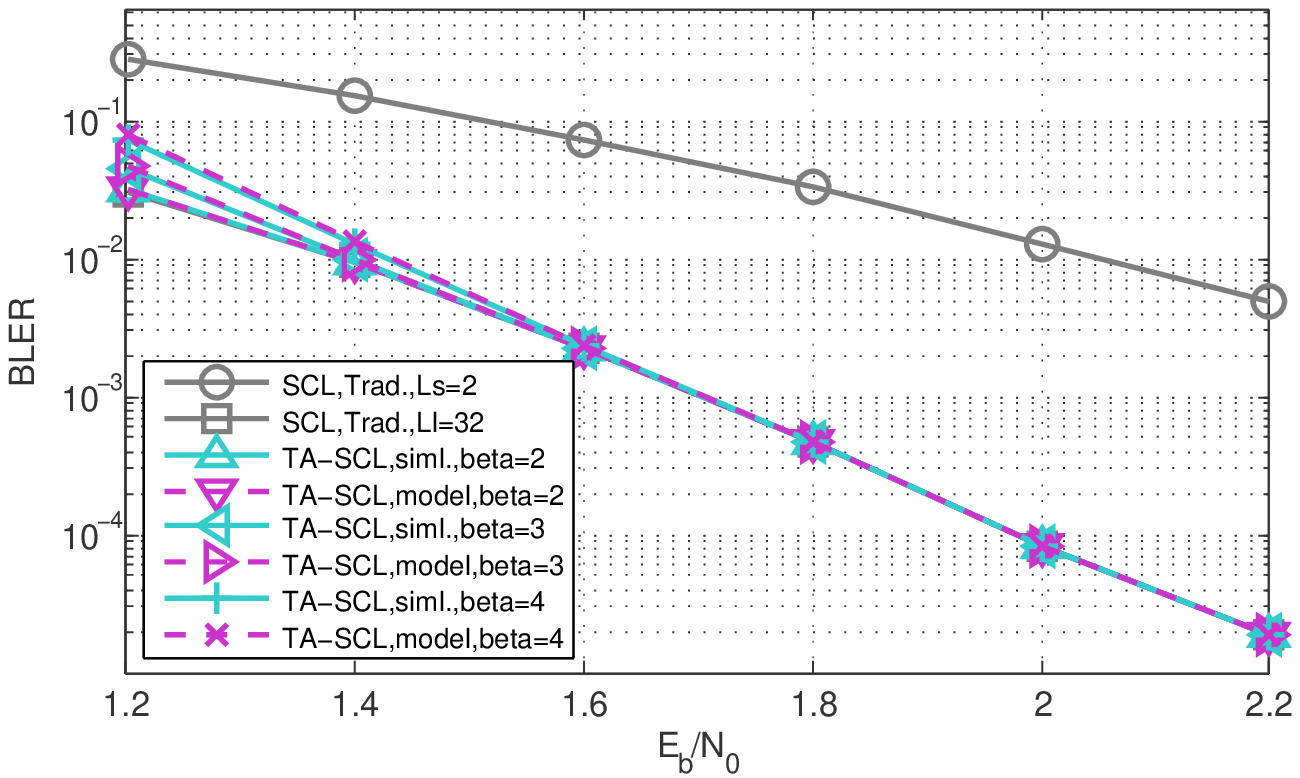}\label{fig:err_perf_beta}}\subfloat[{$\mathbf{D}_{\text{TA}}(3,\zeta)$ ($\zeta\in[1,3]$) with $\mathcal{L}_{s}$=2.}]{\includegraphics[width=6cm]{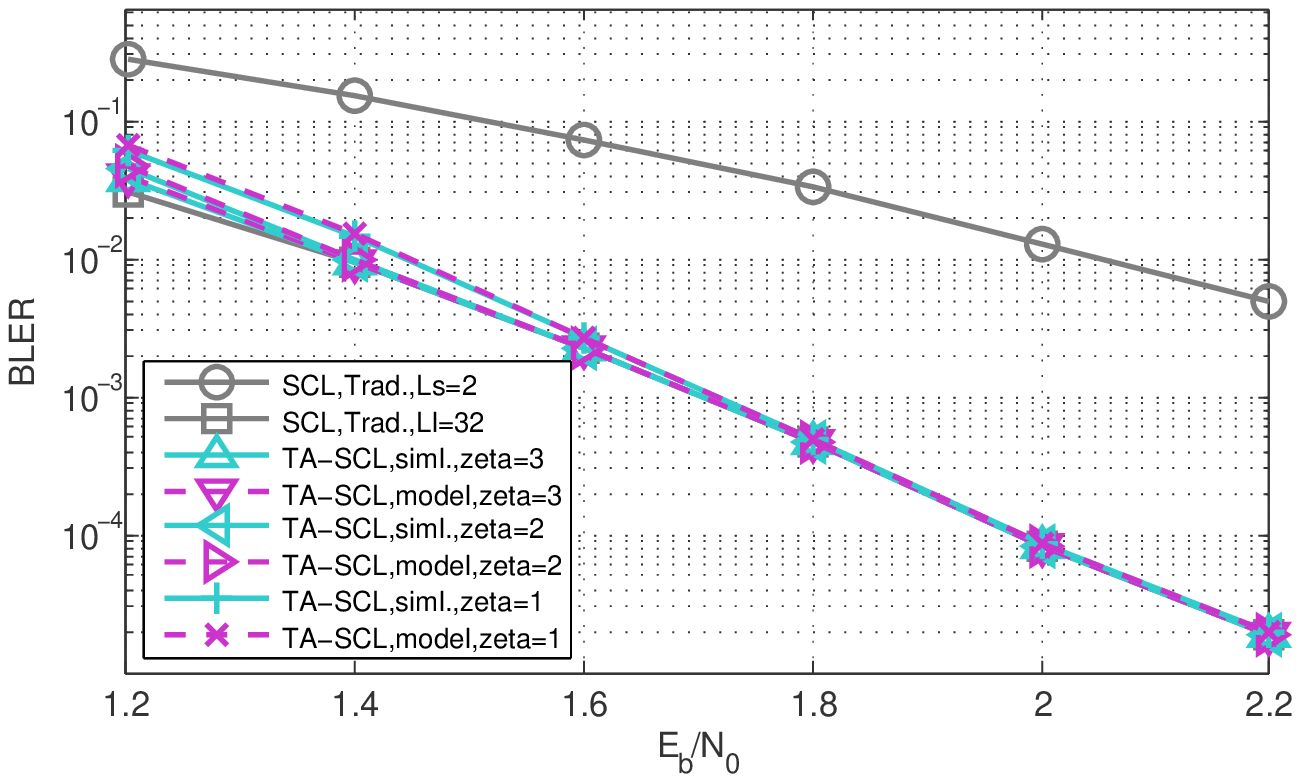}\label{fig:err_perf_zeta}}

\caption{The impact of parameters $\epsilon_{s}$, $\beta$ and $\zeta$ on
the error correction performance of $\mathbf{D}_{\text{TA}}$.}
\label{fig:err_perf}
\end{figure*}
In this sub-section, we study the relationships between the error
correction performance of TA-SCL decoding and its design parameters
based on the proposed model. Simulation results are presented to verify
the analysis numerically. A polar code of $(N,K,r)$=$(1024,512,24)$
is used for simulation over an AWGN channel, which is the same as
that used in \cite{cxia_hkust_tsp_2018_largelist}. The low-latency
hardware-friendly decoding algorithm proposed in \cite{cxia_hkust_tsp_2018_largelist},
multi-bit DTS (MB-DTS), is used for $\textbf{D}_{l}$ with $\mathcal{L}_{l}$=32
as TA-SCL is targeting for hardware-based applications.

Instead of directly study the relationships of the design parameters
with $\epsilon_{\mathbf{D}_{\text{TA}}}$, we first study their relationships
with $\text{Pr(Overflow)}$ with the help of the derived model.
\begin{prop}
If $\beta$ and $\zeta$ are fixed, \textup{$\underset{\epsilon_{s}\to0}{\text{lim}}\text{Pr(Overflow)}=0$}.
\label{prop:err-epsilon}
\end{prop}
\begin{IEEEproof}
As shown in \eqref{eq:trans_mat}, all the elements in a transition
matrix are linear combinations of 1 and $\epsilon_{s}$. Thus, each
element $\pi_{i}$ in the stationary distribution $\pi$ and hence
Pr(Overflow) is a polynomial of $\epsilon_{s}$ and the Proposition
is proved.
\end{IEEEproof}
\prpref \ref{prop:err-epsilon} indicates that TA-SCL decoding should
have a better error correction performance at a higher SNR or if a
larger $\mathcal{L}_{s}$ is used. To verify this, we simulate the
error correction performance of the proposed TA-SCL decoding with
different $\mathcal{L}_{s}$ but the same $\beta$ and $\zeta$ and
the results are shown in \figref \ref{fig:err_perf_es}. We assume
that $\beta$=3, which is a reasonable estimation as will be shown
later in \secref \ref{sec:experiment}. The solid lines show the
simulation results of $\epsilon_{\mathbf{D}_{\text{TA}}}$ and the
dashed lines show the upper bound of $\epsilon_{\mathbf{D}_{\text{TA}}}$
calculated by the proposed model. It can be observed that the TA-SCL
with $\mathcal{L}_{s}$=2 has almost negligible error correction performance
degradation compared to $\mathbf{D}_{l}$ on a wide SNR range. Also,
the performance degradation gradually disappears when SNR increases
and $\epsilon_{s}$ decreases. In contrast, TA-SCL with $\mathcal{L}_{s}$=1
has much poorer performance at low SNR range. This is bacause more
codewords cannot be correctly decoded by $\mathbf{D}_{s}$, and more
operations of $\mathbf{D}_{l}$ are needed, hence $\text{Pr(Overflow)}$
increases. Nevertheless, at high SNR range, the performance degradation
is still negligible. Considering that a smaller $\mathcal{L}_{s}$
usually indicates a lower decoding latency, a larger speed gain can
be achieved. Moreover, it can be observed in \figref \ref{fig:err_perf_es}
that the simulation results are almost the same as the upper bounds
obtained from the analysis, i.e., $\epsilon_{\mathbf{D}_{\text{TA}}}\approx\epsilon_{l}$+$\text{Pr(Overflow)}$.
Thus, the upper bound of the derived model can be used to estimate
the error correction performance of $\mathbf{D}_{\text{TA}}$. %

Next, we study the relationships between $\text{Pr(Overflow)}$ and
the other two design parameters, $\beta$ and $\zeta$.
\begin{itemize}
\item When $\zeta$ increases, more codewords can be stored in the LLR buffer
for $\mathbf{D}_{l}$ decoding. As discussed in \secref \ref{subsec:alg_tascl},
if we have infinity buffer resources, buffer overflow never happens. 
\item When $\beta$ decreases, $\mathbf{D}_{l}$ can decode the codewords
in the LLR buffer sooner after they were stored. When $\beta\leq1$,
$\mathbf{D}_{l}$ decodes a codeword faster than $\mathbf{D}_{s}$
and buffer overflow never happens.
\end{itemize}
\figref \ref{fig:err_perf_beta} and \ref{fig:err_perf_zeta} shows
the performance of TA-SCL with different $\beta$ and $\zeta$, respectively.
It can be seen at low SNR range, both increasing buffer size and decreasing
speed gain $\beta$ lead to a lower $\text{Pr(Overflow)}$ and hence
$\epsilon_{\mathbf{D_{\text{TA}}}}$, which is in accordance with
the discussion above. At high SNR range, the curves of all the $\mathbf{D}_{\text{TA}}(\beta,\zeta)$
are overlapped, which means the error correction performance is good
even when a small buffer is used or a high speed gain is required. 

From the simulation results shown in \figref \ref{fig:err_perf},
the performance loss of TA-SCL decoding $\delta$ is larger at low
SNR range. In practical applications, the decision to select which
$\textbf{D}_{s}$ to use depends on how much performance loss we can
tolerate for a certain BLER. For example, in \figref \ref{fig:err_perf_es},
the performance loss is 20\% at a BLER of $2\cdot10^{-2}$ for $\mathbf{D}_{\text{TA}}(3,3)$
with $\mathcal{L}_{s}$=2 while that for $\mathcal{L}_{s}$=1 is much
larger. Also it can be seen that $\delta$ gradually approaches 0
when SNR increases.

Based on the relationships between the error correction performance
and the design parameters, we summarise the following steps of designing
TA-SCL decoding for a specific polar code.
\begin{enumerate}
\item Running simulations of the target code using $\textbf{D}_{s}$ and
$\textbf{D}_{l}$.
\item Calculating $t_{s}$, $t_{l}$ and corresponding $\beta$. 
\item Gradually increasing the buffer size from $\zeta=1$. Calculating
and checking whether the performance loss $\delta$ is satisfied at
the target BLER by using the Markov model.
\item If $\zeta$ reaches a predefined buffer resource constraint and $\delta$
is still not satisfied, adding idle time to $t_{s}$ to decrease $\beta$
and redo step 3.
\item Running simulations of the target code using the designed $\textbf{D}_{\text{TA}}(\beta,\zeta)$
for verification. %
\end{enumerate}

\section{Hardware Architecture for TA-SCL\label{sec:hw_arc}}

\begin{figure*}
\includegraphics[width=18cm]{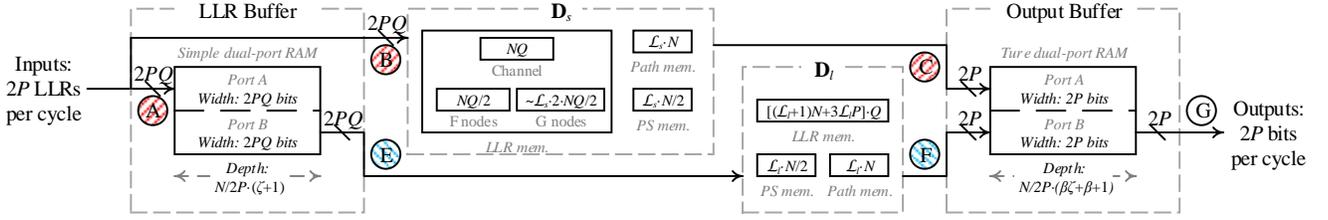}

\caption{Overall hardware architecture and memory usage.}
\vskip 0mm \label{fig:top_arch}
\end{figure*}
\begin{figure*}
\includegraphics[width=18cm]{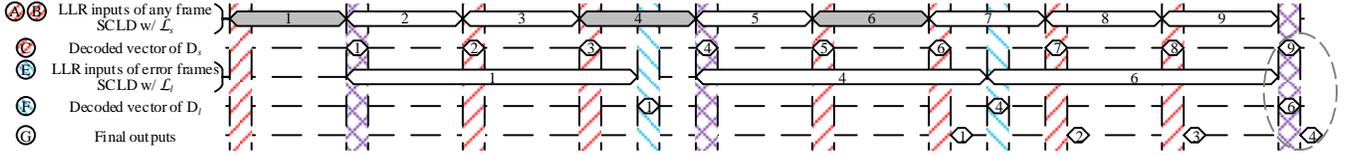}

\caption{Timing schedule of $\mathbf{D}_{\text{TA}}(\frac{5}{2},1)$. The codewords
in gray cannot be decoded correctly by $\mathbf{D}_{s}$.}
\vskip 0mm \label{fig:top_schedule}
\end{figure*}
In this section, we first present the overall architecture of TA-SCL
decoder and analyze its memory usage and timing schedule. Then, we
introduce a low-latency architecture for $\mathbf{D}_{s}$. We denote
the number of clock cycles required for $\mathbf{D}_{s}$ and $\mathbf{D}_{l}$
to decode one frame as $\mathcal{C}_{s}$ and $\mathcal{C}_{l}$,
respectively, then
\begin{equation}
\beta=\frac{t_{l}}{t_{s}}=\frac{\mathcal{C}_{l}}{\mathcal{C}_{s}}.\label{eq:beta_d}
\end{equation}
In the rest of the paper, we will use $\mathcal{C}_{s}$ and $\mathcal{C}_{l}$
to represent the latency instead of $t_{s}$ and $t_{l}$.

\subsection{Overall Architecture \label{subsec:top_arch}}

The proposed architecture of TA-SCL decoder is shown in \figref \ref{fig:top_arch}.
It consists of four major sub-blocks: two constituent SCL decoders
for $\mathbf{D}_{s}$ and $\mathbf{D}_{l}$, an LLR buffer and an
output buffer. Data width of each connection is also marked in \figref
\ref{fig:top_arch}, where $P$ is called parallelism factor, i.e.,
the maximum number of F- or G-nodes that can be executed at the same
time in $\mathbf{D}_{l}$, and $Q$ is the number of quantization
bits for the LLR values. Details of each sub-block is introduced as
below.

To support a high speed gain $\beta$, the architecture of $\mathbf{D}_{s}$
should have a very low decoding latency. Empirically, a $\mathbf{D}_{s}$
with $\mathcal{L}_{s}\leq2$ provides error correction performance
that is good enough to achieve a very low $\text{Pr(Overflow)}$ and
$\epsilon_{\textbf{D}_{\text{TA}}}$, and a larger $\mathcal{L}_{s}$
brings little performance gain but large overhead on timing and hardware
complexity. Hence, a low-latency SCL decoding scheme targeting for
$\mathcal{L}_{s}$=2 is used in the proposed TA-SCL architecture.
It will be introduced in \secref \ref{subsec:ull-scl} in detail.
The memory usage of $\mathbf{D}_{s}$ is shown in \figref \ref{fig:top_arch}.
We only show the major memories which dominate the overall memory
usage. The LLR memory stores both the channel LLRs and the calculated
LLRs during decoding. The channel LLR memory have $NQ$ bits. It is
also re-used for the storage of the calculated LLRs from the G-node
at stage $n$-1. The calculated LLRs from the G-nodes at the other
stages require about $\mathcal{L}_{s}\cdot2\cdot\frac{NQ}{2}$ bits
for storage in total \cite{cxia_hkust_tsp_2018_largelist}. The calculated
LLRs of F-nodes at different stages can share a $\frac{NQ}{2}$-bit
memory as they will be used only once in the next clock cycle directly
and do not need to be stored afterwards \cite{cleroux_mcgill_tsp_2013_semiparallel,cxia_hkust_tsp_2018_largelist}.
The partial-sum memory and path memory requires $\mathcal{L}_{s}\cdot\frac{N}{2}$
bits and $\mathcal{L}_{s}\cdot N$ bits, respectively. This architecture
can also be configured for $\mathcal{L}_{s}$=1, and the corresponding
memory usage is slightly larger than half of that of $\mathbf{D}_{s}$
with $\mathcal{L}_{s}$=2.

For $\mathbf{D}_{l}$, we use the architecture proposed in our previous
work \cite{cxia_hkust_tsp_2018_largelist} and the details will not
be discussed here. The sizes of LLR memory, partial-sum memory and
path memory are $[(\mathcal{L}_{l}+1)N+3\mathcal{L}_{l}P]\cdot Q$
bits, $\mathcal{L}_{l}\cdot\frac{N}{2}$ bits and $\mathcal{L}_{l}\cdot N$
bits, respectively. The number of processing elements (PE) for each
path in $\mathbf{D}_{l}$ equals to $P$.

The LLR buffer is implemented by a simple one-read one-write dual-port
SRAM. The incoming channel LLR values will be written to the channel
LLR memory in $\mathbf{D}_{s}$ and the LLR buffer at the same time.
If $\mathbf{D}_{s}$ could not decode the frame correctly, this frame
data will be kept in the LLR buffer for $\mathbf{D}_{l}$ decoding.
Otherwise, it will be overwritten by the next frame. Thus the total
size of the LLR buffer is $\zeta$+1 frame where $\zeta$ is the number
of buffer required for storing the frame data for $\mathbf{D}_{l}$
as discussed in the analytical model and the additional buffer is
holding the current frame data being decoded by $\mathbf{D}_{s}$.
The size of LLR buffer is hence $NQ\cdot(\zeta$+1) bits. The width
of each port is $2PQ$ bits. This parallelism matches with the I/O
ports of $\mathbf{D}_{l}$ \cite{cxia_hkust_tsp_2018_largelist}. 

The output buffer is implemented by a true dual-port SRAM of which
both ports can be used for reading or writing. It is used to align
the output order to be the same as the input sequence because the
decoding of the frames can be out of order. All the decoding results
from $\mathbf{D}_{s}$ will be temporarily stored in this buffer.
The results will be overwriten if the corresponding codeword is decoded
by $\mathbf{D}_{l}$. Considering the worst case that a TA-SCL reaches
the maximum state, the frame just stored into the LLR buffer and could
not be decoded by $\mathbf{D}_{s}$ correctly will be decoded by $\mathbf{D}_{l}$
after $(\beta\zeta$+$\beta)\cdot\mathcal{C}_{s}$ clock cycles. All
the decoding results of the codewords inputted after this frame need
to be temporarily stored in the output buffer. So the output buffer
needs to accommodate at most $\left\lceil \beta\zeta+\beta\right\rceil $
frames theoretically. In the real design, the output buffer needs
to store $\left\lfloor \beta\zeta+\beta+1\right\rfloor $ frames,
i.e, one more frame of the decoded vectors needs to be stored if $\beta\zeta$+$\beta$
is an integer, and the reason will be explained in \secref \ref{subsec:top_schedule}.
The size of the output buffer is hence $N\cdot\left\lfloor \beta\zeta+\beta+1\right\rfloor $
bits.

The memory usage of TA-SCL is summarized in \tabref \ref{tab:mem_usage}
according to the analysis above. The hardware complexity overhead
of TA-SCL over the traditional architecture for $\mathbf{D}_{l}$
comes from $\mathbf{D}_{s}$ and the two buffers, which is dominated
by the memory overhead. As an example, the memory overhead of $\mathbf{D}_{\text{TA}}(3,3)$
(the one in \figref \ref{fig:err_perf_es}) is also shown in \tabref
\ref{tab:mem_usage} and the overhead is around 20\%. More accurate
experimental results on hardware usage will be presented in \secref
\ref{sec:experiment}. 
\begin{table}
{\small{}\caption{Memory usage of TA-SCL decoding and an example for $\mathbf{D}_{\text{TA}}(3,3)$.
Assuming that $N$=1024, $\mathcal{L}_{l}$=32, $P$=64 and $Q$=6}
\label{tab:mem_usage}}\setlength{\tabcolsep}{2pt}

{\small{}}%
\begin{tabular}{c|c|c|c}
\hline 
{\small{}Sub-blocks} & {\small{}Memory usage (bits)} & {\small{}$\mathcal{L}_{s}$=2} & {\small{}$\mathcal{L}_{s}$=1}\tabularnewline
\hline 
{\small{}$\mathbf{D}_{l}$} & {\small{}$[(\mathcal{L}_{l}$+$1)N$+$3\mathcal{L}_{l}P]\cdot Q$+$\mathcal{L}_{l}\cdot\frac{3N}{2}$} & {\small{}288,768} & {\small{}288,768}\tabularnewline
\hline 
{\small{}Others} & {\small{}$(\zeta$+$\mathcal{L}_{s}$+$\frac{5}{2}$)$NQ$+$(\left\lfloor \beta\zeta\right.$+$\beta$+$\left.1\right\rfloor $+$\frac{3\mathcal{L}_{s}}{2}$)$N$} & {\small{}62,464} & {\small{}54,784}\tabularnewline
\hline 
\multicolumn{2}{c|}{{\small{}Overhead}} & {\small{}21\%} & {\small{}19\%}\tabularnewline
\hline 
\end{tabular}\vskip -5mm 
\end{table}

\subsection{Timing schedule \label{subsec:top_schedule}}

The timing schedule of TA-SCL architecture is illustrated in \figref
\ref{fig:top_schedule} with an example for $\mathbf{D}_{\text{TA}}(\frac{5}{2},1)$.
Each line represents a decoding process or a data flow in \figref
\ref{fig:top_arch}, as marked with circled letters. The number shown
in the waveforms represent the frame indices. 

The first and third rows represent the decoding operations of $\mathbf{D}_{s}$
and $\mathbf{D}_{l}$, respectively, which are similar as the timing
schedule shown in \figref \ref{fig:schedule}. The first $\mathcal{C}_{\text{rw}}\triangleq N/2P$
clock cycles are used to load the input channel LLRs. The periods
filled with ``//'', ``\textbackslash{}\textbackslash{}'' and ``X''
stripes represent the LLR loading time for $\mathbf{D}_{s}$, for
$\mathbf{D}_{l}$ and for both, respectively. The second and fourth
rows represent the operations of sending the decoding results from
$\mathbf{D}_{s}$ and $\mathbf{D}_{l}$ to the output buffer, respectively.
These two operations are executed concurrectly with the loading operations
of the next codeword to be decoded. The fifth row shows when the final
output results are generated from the output buffer. From the timing
schedule, it can be seen that the final decoding results of any codeword
will be available after $\mathcal{C}_{s}$+$\mathcal{C}_{s}$$\cdot(\beta\zeta$+$\beta$)+$\mathcal{C}_{\text{rw}}$
clock cycles when the corresponding LLRs are inputted to the decoder.
The first term is caused by $\mathbf{D}_{s}$. The second term is
dictated by the worst case discussed in \secref \ref{subsec:top_arch}.
The third term is added to avoid potential memory collision as circled
in \figref \ref{fig:top_schedule}: the decoded data is read out
from the output buffer only after the two decoders send the decoded
results to the output buffer. This is also the reason why we need
space for one more frame as mentioned in \secref \ref{subsec:top_arch}.
As an example, the system latency of $\mathbf{D}_{\text{TA}}(\frac{5}{2},1)$
in \figref \ref{fig:top_arch} is $6\mathcal{C}_{s}$+$\mathcal{C}_{\text{rw}}$
clock cycles.

\subsection{Low-Latency SCL Decoding Scheme\label{subsec:ull-scl}}

\begin{figure}
\includegraphics[width=8.8cm]{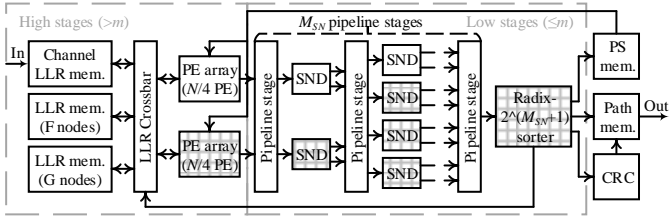}

\caption{Hardware architecture of $\textbf{D}_{s}$. The blocks with ``+''
stripes are disabled when $\mathcal{L}_{s}$=1.}
\label{fig:ds_arch}
\end{figure}
\begin{table}
{\small{}\caption{Summarization of special nodes used at low stages}
\label{tab:special_nodes}}\setlength{\tabcolsep}{1mm}

{\small{}}%
\begin{tabular}{c|c|c||c|c|c||c|c|c}
\hline 
{\small{}Name} & {\small{}\# frz.} & {\small{}\# inf.} & {\small{}Name} & {\small{}\# frz.} & {\small{}\# inf.} & {\small{}Name} & {\small{}\# frz.} & {\small{}\# inf.}\tabularnewline
\hline 
{\small{}Rate-0} & {\small{}$T$} & {\small{}$0$} & {\small{}Rep.} & {\small{}$T$-1} & {\small{}1} & {\small{}Rep2} & {\small{}$T$-2} & {\small{}2}\tabularnewline[0.05cm]
\hline 
{\small{}Rate-1} & {\small{}0} & {\small{}$T$} & {\small{}SPC} & {\small{}1} & {\small{}$T$-1} & {\small{}SPC2} & {\small{}2} & {\small{}$T$-2}\tabularnewline[0.05cm]
\hline 
\end{tabular}\vskip -3mm 
\end{table}
\begin{table*}
{\small{}\caption{Code settings, numbers of 16-bit sub-codes in different groups and
latency of the two component decoders}
\label{tab:code_eg}}\setlength{\tabcolsep}{1mm}

{\small{}}%
\begin{tabular}{c|c|c|c|c||c|c|c|c||>{\centering}p{1cm}|>{\centering}p{1cm}|>{\centering}p{1cm}||c|c|c|c|c}
\hline 
\multirow{2}{*}{{\small{}Codes}} & \multicolumn{4}{c||}{{\small{}I. Code Settings}} & \multicolumn{4}{c||}{{\small{}II. \# of sub-codes with $\mathcal{F}_{i}=$ }} & \multicolumn{3}{c||}{{\small{}III. Latency of $\mathbf{D}_{s}$ ($\mathcal{L}_{s}$=2)}} & \multicolumn{5}{c}{{\small{}IV. Latency of $\mathbf{D}_{l}$}}\tabularnewline
\cline{2-17} 
 & {\small{}$N$} & {\small{}$K$} & {\small{}$r$} & {\small{}$|\mathcal{A}_{r}|$} & {\small{}0,16 (1)} & {\small{}1,2,14,15 (2)} & {\small{}7,8,9 (3)} & {\small{}others (4)} & {\small{}$\mathcal{C}_{\text{MBD}}$} & {\small{}$\mathcal{C}_{\text{SCD}}$} & {\small{}$\mathcal{C}_{s}$} & {\small{}$\mathcal{C}_{\text{LM}}$} & {\small{}$\mathcal{C}_{\text{SCD}}$} & {\small{}$\mathcal{C}_{\text{fine}}$} & {\small{}$\mathcal{C}_{\text{zero}}$} & {\small{}$\mathcal{C}_{l}$}\tabularnewline
\hline 
\hline 
{\small{}$\mathcal{P}_{1}$} & {\small{}1024} & {\small{}512} & {\small{}24} & {\small{}360} & {\small{}32} & {\small{}13} & {\small{}2} & {\small{}17} & {\small{}132} & {\small{}63} & {\small{}203} & {\small{}520} & {\small{}296} & {\small{}-128} & {\small{}-41} & {\small{}647}\tabularnewline
\hline 
{\small{}$\mathcal{P}_{2}$} & {\small{}1024} & {\small{}768} & {\small{}24} & {\small{}622} & {\small{}38} & {\small{}12} & {\small{}2} & {\small{}12} & {\small{}116} & {\small{}63} & {\small{}187} & {\small{}501} & {\small{}296} & {\small{}-128} & {\small{}-18} & {\small{}651}\tabularnewline
\hline 
{\small{}$\mathcal{P}_{3}$} & {\small{}256} & {\small{}128} & {\small{}8} & {\small{}78} & {\small{}6} & {\small{}4} & {\small{}2} & {\small{}4} & {\small{}36} & {\small{}15} & {\small{}53} & {\small{}152} & {\small{}66} & {\small{}-32} & {\small{}-18} & {\small{}168}\tabularnewline
\hline 
\end{tabular}\vskip -4mm 
\end{table*}

In this section, we introduce a low-latency SCL (LL-SCL) decoding
scheme customized for $\mathbf{D}_{s}$ with $\mathcal{L}_{s}\leq$2
such that $\mathbf{D}_{s}$ can support a large speed gain. It combines
several state-of-the-art low-latency decoding schemes for SCL decoding,
including G-node look-ahead scheme \cite{czhang_umn_icc_2012_lookahead},
multi-bit decoding (MBD) \cite{crxiong_lehigh_tvlsi_2016_sclmm} and
special node decoding (SND) \cite{sahashemi_mcgill_tsp_2017_fastflexible,sahashemi_mcgill_tcasi_2016_ssclspc}.
Specifically, LL-SCL can be divided into the following two parts.
\begin{itemize}
\item SC calculations at stages not lower than stage $m$=$\log M_{s}$
($M_{s}$ is the number of merged bits for MBD in $\mathbf{D}_{s}$)
are calculated by normal SC algorithm with the full parallelism, i.e.,
$\frac{N}{2}$ PEs are used and any node in the scheduling tree takes
only one clock cycle. Moreover, G-node look-ahead scheme \cite{czhang_umn_icc_2012_lookahead}
is used so that each pair of sibling nodes are calculated simultaneously
and half of the latency is saved. Thus, the decoding latency of this
part is $\frac{N}{M_{s}}$-1 clock cycles. 
\item SC calculations at the low stages are replaced with MBD which decodes
$M_{s}$ bits in the same sub-tree rooted at stage $m$ simultaneously.
According to \cite{crxiong_lehigh_tvlsi_2016_sclmm}, given a certain
number of frozen bit, there is only one code pattern for the $M_{s}$-bit
sub-codes. So totally there are only $M_{s}$+1 different code patterns.
To reduce the decoding latency, the decoding scheme for each code
pattern is designed as follows. These $M_{s}$+1 code patterns are
divided into multiple special nodes as shown in \tabref \ref{tab:special_nodes}
where $T$ is the number of bits in each special node. The corresponding
numbers of information bits and frozen bits are also listed. Rate-0,
rate-1, repetition (Rep.) and single parity check (SPC) nodes are
decoded according to the schemes presented in \cite{gsarkis_mcgill_jsac_2016_sclfast,sahashemi_mcgill_tsp_2017_fastflexible,sahashemi_mcgill_tcasi_2016_ssclspc}.
Rep2/SPC2 nodes can be divided into two Rep./SPC nodes with half of
the lengths which can be decoded concurrently. Thus, each special
node can be decoded by SND within one clock cycle. The number of paths
is doubled after a special node is decoded. Different from the traditional
SCL decoding, all the expanded paths are kept temporarily. List pruning
is not executed until the end of each $M_{s}$-bit sub-code, which
takes another clock cycle. Thus, an $M_{s}$-bit sub-code that can
be divided into $M_{\text{SN}}$ special nodes requires $M_{\text{SN}}$+1
clock cycles to decode, except for $M_{s}$-bit rate-0 and rate-1
nodes which do not need a sorting operation and only take 1 clock
cycle.
\end{itemize}
The total decoding latency of the LL-SCL decoding scheme is
\begin{align}
\mathcal{C}_{s} & =\mathcal{C}_{\text{MBD}}+\mathcal{C}_{\text{SCD}}+\mathcal{C}_{\text{rw}},\label{eq:lat_ta_scl}\\
 & =\sum_{i=1}^{N/M_{s}}(M_{\text{SN}}(\mathcal{F}_{i})+\mathcal{C}_{\text{sort}}(\mathcal{F}_{i}))+(\frac{N}{M_{s}}-1)+\frac{N}{2P},\label{eq:lat_detail}
\end{align}
where $\mathcal{C}_{\text{rw}}$ is the LLR loading latency and $\mathcal{F}_{i}$
is the number of frozen bits in the $i^{th}$ $M_{s}$-bit sub-code,
$M_{\text{SN}}(\mathcal{F}_{i})$ and $\mathcal{C}_{\text{sort}}(\mathcal{F}_{i})$
are the decoding clock cycles for SND and sorting of the code pattern
corresponding to $\mathcal{F}_{i}$, respectively. For $\mathcal{L}_{s}$=2,
\begin{equation}
\mathcal{C}_{\text{sort}}(\mathcal{F}_{i})=\begin{cases}
0, & \text{if }\mathcal{F}_{i}=0\text{ or }M_{s},\\
1, & \text{otherwise}.
\end{cases}\label{eq:c_sort}
\end{equation}
For $\mathcal{L}_{s}$=1, the sorting operation for list pruning is
unnecessary. Hence, $\mathcal{C}_{\text{sort}}(\mathcal{F}_{i})$=0
for $\mathcal{L}_{s}$=1, which indicates the throughput can be increased
by using conventional SC decoding as $\mathbf{D}_{s}$. 

The top level architecture of LL-SCL is shown in \figref \ref{fig:ds_arch}.
For high stages, the SCL architecture \cite{abalatsoukas_epfl_tsp_2015_llrlscd}
uses a PE array with $\frac{N}{4}$ PEs for each path. As the calculations
at the highest stage $n$-1, which requires $\frac{N}{2}$ PEs to
compute, are the same for both paths, they can use the two PE arrays
from the two paths to accomplish. For low stages, $\text{max}(M_{\text{SN}})$
stages of SND blocks and a radix-$2^{\text{max}(M_{\text{SN}})+1}$
sorter are implemented in a feedforward manner, which are directly
mapped from the decoding schemes introduced above. The architecture
can be configured for $\mathcal{L}_{s}$=1 by simply disabling some
of the blocks as shown in \figref \ref{fig:ds_arch}. 

\section{Experimental Results\label{sec:experiment}}

\subsection{Decoding Latency and Error Correction Performance of TA-SCL Decoding
for Practical Polar Codes\label{subsec:code_eg}}

\begin{table}
\caption{Different TA-SCL decoders for $\mathcal{P}_{1}$\textasciitilde{}$\mathcal{P}_{3}$}
\label{tab:impl_dec}

\begin{tabular}{c|c||c|c||c|c||c|c}
\hline 
{\small{}$\mathbf{D}_{\text{TA}}$} & {\small{}Code} & {\small{}$\mathcal{L}_{s}$} & {\small{}$\mathcal{C}_{s}$} & {\small{}$\mathcal{L}_{l}$} & {\small{}$\mathcal{C}_{l}$} & {\small{}$\text{\ensuremath{\beta}}$} & {\small{}$\zeta$}\tabularnewline
\hline 
\hline 
{\small{}$\mathbf{D}_{1}$} & {\small{}$\mathcal{P}_{1}$} & {\small{}2} & {\small{}203} & {\small{}32} & {\small{}647} & {\small{}3.18} & {\small{}2}\tabularnewline
\hline 
{\small{}$\mathbf{D}_{2}$} & {\small{}$\mathcal{P}_{2}$} & {\small{}2} & {\small{}187} & {\small{}32} & {\small{}651} & {\small{}3.48} & {\small{}2}\tabularnewline
\hline 
{\small{}$\mathbf{D}_{3}$} & {\small{}$\mathcal{P}_{3}$} & {\small{}2} & {\small{}53} & {\small{}32} & {\small{}168} & {\small{}3.17} & {\small{}2}\tabularnewline
\hline 
{\small{}$\mathbf{D}_{4}$} & {\small{}$\mathcal{P}_{1}$} & {\small{}1} & {\small{}170} & {\small{}32} & {\small{}647} & {\small{}3.80} & {\small{}6}\tabularnewline
\hline 
{\small{}$\mathbf{D}_{5}$} & {\small{}$\mathcal{P}_{1}$} & {\small{}2} & {\small{}203} & {\small{}8} & {\small{}647} & {\small{}3.18} & {\small{}2}\tabularnewline
\hline 
\end{tabular}
\end{table}
\begin{figure*}
\includegraphics{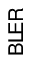}\subfloat[$\mathbf{D}_{1}$ for $\mathcal{P}_{1}$.]{\includegraphics[width=2.9cm]{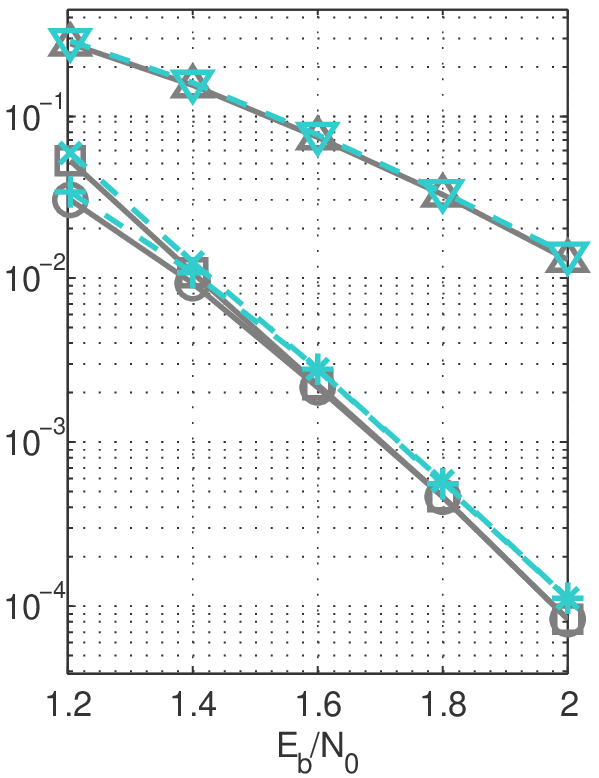}

\label{fig:tascl_p-1}}\subfloat[$\mathbf{D}_{2}$ for $\mathcal{P}_{2}$.]{\includegraphics[width=2.9cm]{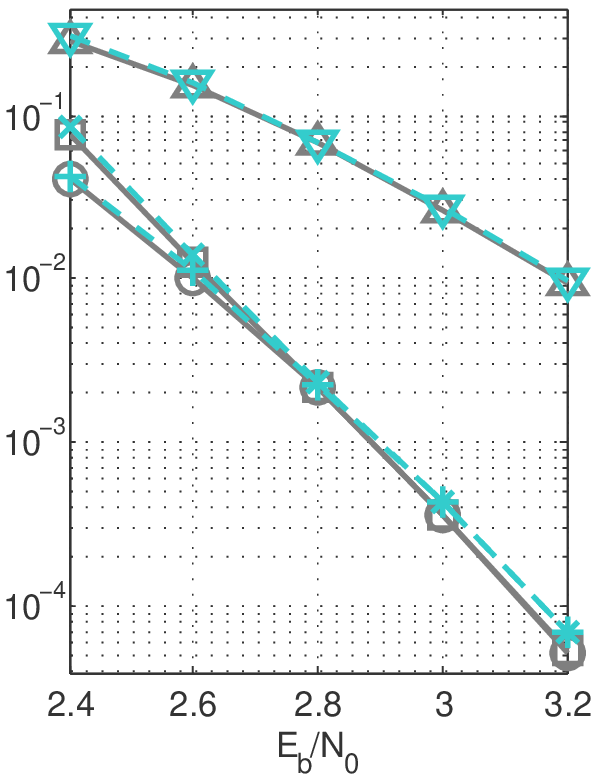}\label{fig:tascl_p-2}}\subfloat[$\mathbf{D}_{3}$ for $\mathcal{P}_{3}$.]{\includegraphics[width=2.9cm]{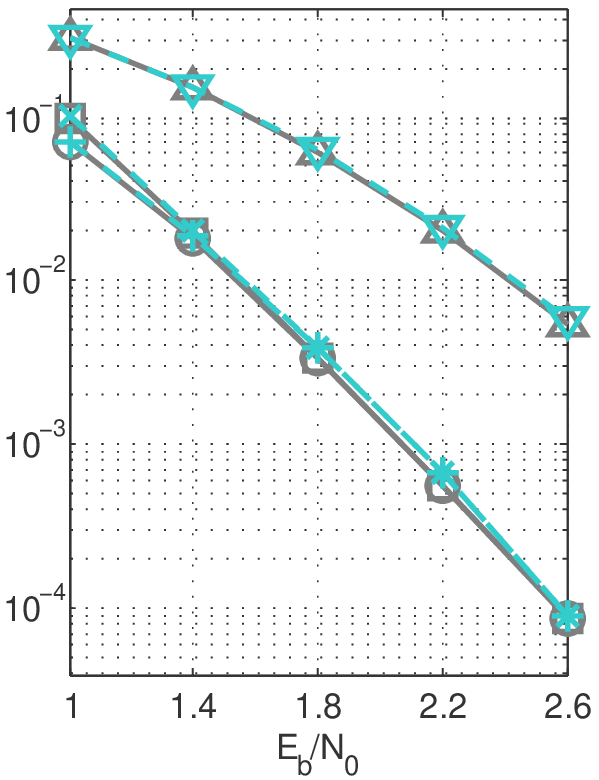}\label{fig:tascl_p-3}}\subfloat[$\mathbf{D}_{4}$ for $\mathcal{P}_{1}$.]{\includegraphics[width=2.9cm]{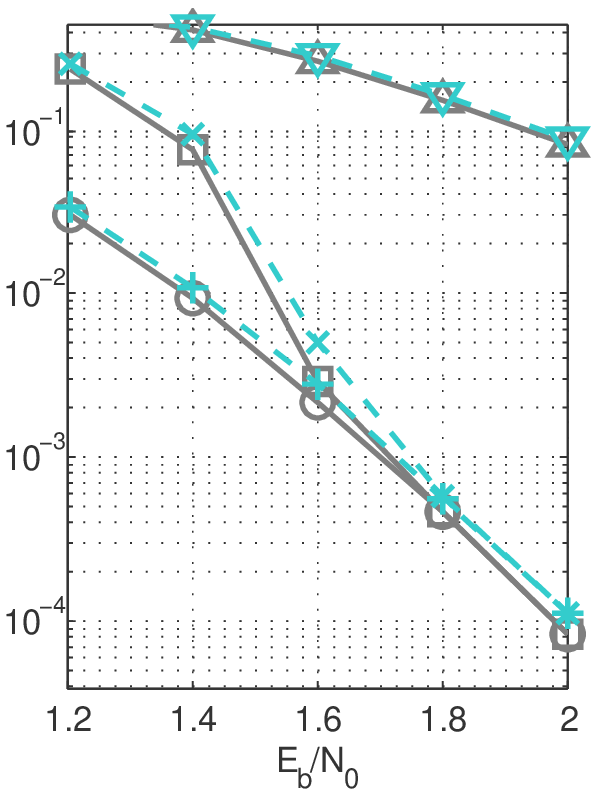}\label{fig:tascl_p-4}}\subfloat[$\mathbf{D}_{5}$ for $\mathcal{P}_{1}$.]{\includegraphics[width=2.9cm]{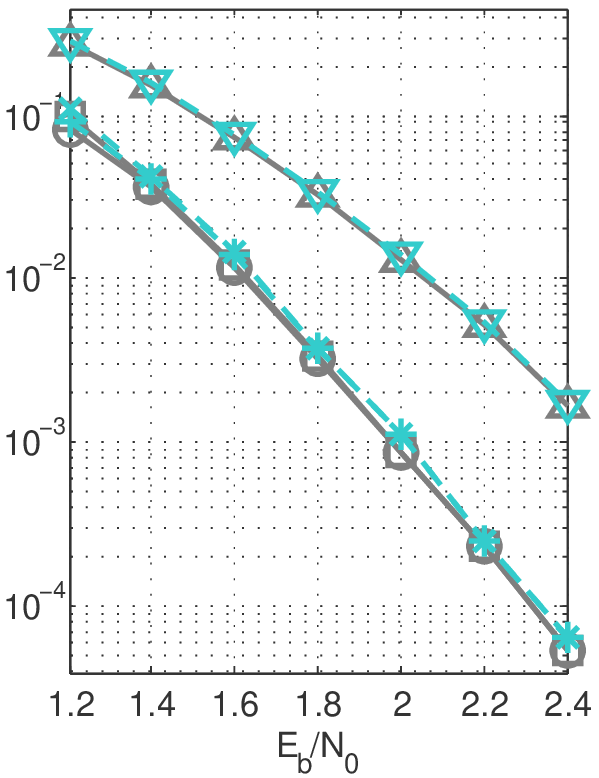}

\label{fig:tascl_p-5}}

\includegraphics[width=18cm]{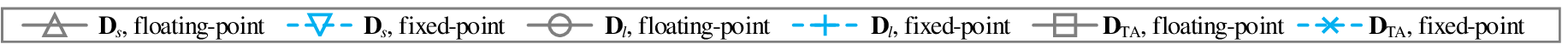}\caption{Error correction performance of TA-SCL decoders $\mathbf{D}_{1}$\textasciitilde{}$\mathbf{D}_{5}$
shown in \tabref \ref{tab:impl_dec}.}
\end{figure*}
\begin{table*}
\caption{Synthesis results of the proposed TA-SCL decoders and comparison with
state-of-the-art SCL decoders for 1024-bit polar codes}
\label{tab:syn_result}\setlength{\tabcolsep}{1mm}

{\small{}}%
\begin{tabular}{>{\centering}p{4cm}||c|c|c|c|c|c|c|>{\raggedright}m{5cm}}
\hline 
 & {\small{}$\mathbf{D}_{1}$} & {\small{}$\mathbf{D}_{4}$} & {\small{}$\mathbf{D}_{5}$} & \multicolumn{2}{c|}{{\small{}\cite{cxia_hkust_tsp_2018_largelist} ($M_{l}$=8)}} & {\small{}$\,$\cite{sahashemi_mcgill_tsp_2017_fastflexible}$\dagger$} & {\small{}$\,$\cite{jlin_lehigh_tvlsi_2016_highthpt}$\diamond$ } & \tabularnewline
\cline{1-8} 
{\small{}$K=|\mathcal{A}|$} & {\small{}512} & {\small{}512} & {\small{}512} & {\small{}512} & {\small{}512} & {\small{}528} & {\small{}512} & {\footnotesize{}Notes: }\tabularnewline
\cline{1-8} 
{\small{}List size $\mathcal{L}$} & {\small{}32} & {\small{}32} & {\small{}8} & {\small{}32} & {\small{}8} & {\small{}8} & {\small{}8} & \multirow{3}{5cm}{{\footnotesize{}$\dagger$ The synthesis results in \cite{sahashemi_mcgill_tsp_2017_fastflexible}
are based on TSMC 65nm technology and are scaled to a 90nm technology. }}\tabularnewline
\cline{1-8} 
{\small{}Clock freq. (MHz)} & {\small{}465} & {\small{}465} & {\small{}595} & {\small{}417} & {\small{}556} & {\small{}520} & {\small{}289} & \tabularnewline
\cline{1-8} 
{\small{}Throughput (Mbps)} & {\small{}2346} & {\small{}2801} & {\small{}3002} & {\small{}827} & {\small{}1103} & {\small{}862} & {\small{}732} & \tabularnewline
\cline{1-8} 
{\small{}Total area ($\text{mm}^{2}$)} & {\small{}22.00} & {\small{}21.27} & {\small{}7.67} & {\small{}19.58} & {\small{}4.54} & {\small{}7.64} & {\small{}7.22} & \multirow{2}{5cm}{{\footnotesize{}$\diamond$ The synthesis results in \cite{jlin_lehigh_tvlsi_2016_highthpt}
are based on TSMC 90nm technology.}}\tabularnewline
\cline{1-8} 
{\small{}Area efficiency (Mbps/$\text{mm}^{2}$)} & {\small{}106.63} & {\small{}131.69} & {\small{}391.40} & {\small{}42.34} & {\small{}242.95} & {\small{}112.83} & {\small{}101.36} & \tabularnewline
\hline 
\end{tabular}\vskip -2mm 
\end{table*}
\begin{table}
\caption{Area breakdown of the proposed TA-SCL decoders}
\label{tab:area_breakdown}\setlength{\tabcolsep}{0.8mm}%

{\small{}}%
\begin{tabular}{>{\raggedright}p{1cm}|c|cc|cc|cc}
\hline 
\multicolumn{2}{c|}{} & \multicolumn{2}{c|}{{\small{}$\mathbf{D}_{1}$}} & \multicolumn{2}{c|}{{\small{}$\mathbf{D}_{4}$}} & \multicolumn{2}{c}{{\small{}$\mathbf{D}_{5}$}}\tabularnewline
\hline 
\hline 
\multirow{5}{1cm}{{\small{}Area ($\text{mm}^{2}$)}} & {\small{}$\mathbf{D}_{s}$} & {\small{}1.97} & {\small{}(9\%)} & {\small{}1.07} & {\small{}(5\%)} & {\small{}1.97} & {\small{}(25\%)}\tabularnewline
\cline{2-8} 
 & {\small{}$\mathbf{D}_{l}$ ($M_{l}$=4)} & {\small{}18.76} & {\small{}(85\%)} & {\small{}18.76} & {\small{}(88\%)} & {\small{}4.43} & {\small{}(58\%)}\tabularnewline
\cline{2-8} 
 & {\small{}LLR buffer} & {\small{}1.14} & {\small{}(5\%)} & {\small{}1.24} & {\small{}(6\%)} & {\small{}1.14} & {\small{}(15\%)}\tabularnewline
\cline{2-8} 
 & {\small{}Output buffer} & {\small{}0.13} & {\small{}(1\%)} & {\small{}0.20} & {\small{}(1\%)} & {\small{}0.13} & {\small{}(2\%)}\tabularnewline
\cline{2-8} 
 & {\small{}Total area} & {\small{}22.00} & {\small{}(100\%)} & {\small{}21.27} & {\small{}(100\%)} & {\small{}7.67} & {\small{}(100\%)}\tabularnewline
\hline 
\end{tabular}\vskip -3mm 
\end{table}
In this sub-section, we demonstrate the speed gain and the error correction
performance of the proposed architectures. We apply TA-SCL decoding
on several polar codes with different code lengths and code rates
as shown in \tabref \ref{tab:code_eg} for illustration. These codes
are similar to those chosen for the 5G eMBB control channel \cite{3gpp_3gpp_ran087_2016_5g}.
$\mathcal{P}_{1}$ is the same as the one used in \cite{cxia_hkust_tsp_2018_largelist}
and similar to those presented in many existing works \cite{sahashemi_mcgill_tsp_2017_fastflexible,jlin_lehigh_tvlsi_2016_highthpt}.
$\mathcal{P}_{2}$ and $\mathcal{P}_{3}$ have different code rate
and code length, respectively, and are used as examples to show the
flexibility of TA-SCL decoding.

\tabref \ref{tab:code_eg} summarizes code characteristics and the
number of cycles required for the corresponding $\mathbf{D}_{s}$
and $\mathbf{D}_{l}$. $|\mathcal{A}_{r}|$ is the number of reliable
bits for the MB-DTS in the corresponding $\mathbf{D}_{l}$ \cite{cxia_hkust_tsp_2018_largelist}.
For $\mathbf{D}_{s}$, MBD is applied to each 16-bit sub-codes, i.e.,
$M_{s}$=16, which means there are 17 code patterns in total. The
number of special nodes in each code pattern and hence the number
of cycles required for decoding is
\begin{equation}
M_{\text{SN}}(\mathcal{F}_{i})=\begin{cases}
1, & \text{if }\mathcal{F}_{i}=0,1,2,14,15,16,\\
2, & \text{if }\mathcal{F}_{i}=7,8,9,\\
3, & \text{if }\mathcal{F}_{i}=3,4,5,6,10,11,12,13,
\end{cases}\label{eq:msn_16}
\end{equation}
As $\text{max}(M_{\text{SN}})$=3, the decoding latency of a sub-code
varies from one to four clock cycles according to \eqref{eq:c_sort}
and \eqref{eq:msn_16}. As shown in part II of \tabref \ref{tab:code_eg},
these code patterns are divided into four different groups according
to their decoding latency as shown in the brackets. The numbers of
sub-codes in each group are then shown in the table, with which the
decoding latency of $\mathbf{D}_{s}$ can be calculated according
to \eqref{eq:lat_detail} and presented in part III. For $\mathbf{D}_{l}$,
MB-DTS is applied to sub-codes with a maximum length of $M_{l}$=4
and $P$=64 PEs are used for each path. Its total latency is calculated
according to\cite{cxia_hkust_tsp_2018_largelist} and the results
are shown in part IV. With the selected $M_{s}$ and $M_{l}$, the
critical path delays of $\mathbf{D}_{s}$ and $\mathbf{D}_{l}$ are
similar (both require 4\textasciitilde{}5 stages of adders delay)
and the decoding latency can be minimized.

Based on the parameters shown in \tabref \ref{tab:code_eg}, we can
design the TA-SCL decoders for $\mathcal{P}_{1}$\textasciitilde{}$\mathcal{P}_{3}$
to meet different requirements of error correction performance and
speed gain. Some designs are listed in \tabref \ref{tab:impl_dec}.
If we want to target for a TA-SCL decoder that has good decoding performance
across a wide SNR range, as shown in \figref \ref{fig:err_perf},
$\mathcal{L}_{s}$=2 should be used. $\mathbf{D}_{1}$, $\mathbf{D}_{2}$
and $\mathbf{D}_{3}$ shown in \tabref \ref{tab:impl_dec} are the
designs for $\mathcal{P}_{1}$, $\mathcal{P}_{2}$ and $\mathcal{P}_{3}$,
respectively, with $\mathcal{L}_{s}$=2 and $\mathcal{L}_{l}$=32.
To determine the values of $\beta$ and $\zeta$ for each design,
we set the maximum performance loss $\delta$ at a BLER of $10^{-2}$
to be 30\%. $\beta$ is then calculated according to \eqref{eq:beta_d}
and $\zeta$ is obtained by using the design flow presented in \secref
\ref{subsec:err_perf_tascl}. It can be observed from \tabref \ref{tab:impl_dec}
that the speed gain is 3x\textasciitilde{}3.5x for these decoders.
The error correction performance of these decoders on AWGN channel
are simulated using both floating-point and fixed-point numbers and
the results are shown in \figref \ref{fig:tascl_p-1}-\ref{fig:tascl_p-3}.
The quantization schemes for $\mathbf{D}_{s}$ are $Q_{s,\text{LLR}}$=7
and $Q_{s,\text{PM}}$=8 so that the performance loss due to quantization
error is reduced to the minimum and this is essential to avoid performance
loss for the TA-SCL decoding as analysed in \secref \ref{sec:ta-scl}.
The quantization schemes for $\mathbf{D}_{l}$ are $Q_{l,\text{LLR}}$=6
and $Q_{l,\text{PM}}$=8 to achieve a balance between performance
loss and hardware complexity. It can be seen that the fixed-point
simulation results of TA-SCL decoding has negligible performance degradation
(\textless{}0.05dB) compared with the floating-point results of $\mathbf{D}_{l}$
at the BLER of $10^{-2}$.

For some applications that only need to work at high SNR range but
require a high throughput, we can use a $\mathbf{D}_{\text{TA}}$
with $\mathcal{L}_{s}$=1. We illustrate the performance of such a
decoder with an example of $\mathbf{D}_{4}$ shown in \tabref \ref{tab:impl_dec}
for $\mathcal{P}_{1}$. The decoding latency of $\mathbf{D}_{s}$
with $\mathcal{L}_{s}$=1 is 32 clock cycles fewer than that of $\mathbf{D}_{s}$
with $\mathcal{L}_{s}$=2. Consequently, $\mathbf{D}_{4}$ can achieve
0.6x more speed gain compared with $\mathbf{D}_{1}$. The error correction
performance of $\mathbf{D}_{4}$ is shown in \figref \ref{fig:tascl_p-4}.
As $\epsilon_{s}$ and hence Pr(Overflow) is large at a lower SNR,
we set the maximum performance loss $\delta$ at a lower BLER, i.e.,
a higher SNR. In this case, $\delta$ is set to be 30\% at a BLER
of $10^{-3}$ and the LLR buffer size $\zeta$ is equal to 6 to achieve
this performance .

For some applications that can trade the error correction performance
with lower decoding complexity, $\mathcal{L}_{l}$ can be smaller
than 32 \cite{cxia_hkust_tsp_2018_largelist}. $\mathbf{D}_{5}$ shown
in \tabref \ref{tab:impl_dec} is an example for $\mathcal{P}_{1}$.
This design has the same design parameters, $\beta$ and $\zeta$,
as those of $\mathbf{D}_{1}$. The hardware complexity of $\mathbf{D}_{5}$
is much smaller than that of $\mathbf{D}_{1}$ and the results will
be shown later. The error correction performance of $\mathbf{D}_{5}$
is shown in \figref \ref{fig:tascl_p-5}. 

The simulation results show that by just using a smaller number of
buffer, e.g. $\zeta$=2, the maximum speed gain $\beta$ can be achieved
for all the designs. To show the actual throughput gain achieved by
TA-SCL on hardware, we realize the design of the TA-SCL decoders $\mathbf{D}_{1}$,
$\mathbf{D}_{4}$ and $\mathbf{D}_{5}$ for $\mathcal{P}_{1}$ and
obtain their throughputs. The results will be presented in the next
sub-section and compared with the results of the state-of-the-art
polar decoders \cite{cxia_hkust_tsp_2018_largelist,sahashemi_mcgill_tsp_2017_fastflexible,jlin_lehigh_tvlsi_2016_highthpt}.

\subsection{Implementation Results of the Proposed Architecture for TA-SCL Decoding\label{subsec:implementation}}

The proposed architecture is synthesized with a UMC 90nm CMOS process
using Synopsys Design Compiler. The quantization schemes and the number
of PEs for $\mathbf{D}_{l}$ are the same as those presented in \cite{cxia_hkust_tsp_2018_largelist,sahashemi_mcgill_tsp_2017_fastflexible,jlin_lehigh_tvlsi_2016_highthpt}
for a fair comparison. The reported throughputs are in terms of coded
bits and the reported area includes both cell and net area. 

\tabref \ref{tab:syn_result} summarizes the synthesis results of
the TA-SCL decoder $\mathbf{D}_{1}$, $\mathbf{D}_{4}$ and $\mathbf{D}_{5}$
for the polar code $\mathcal{P}_{1}$. The results of \cite{cxia_hkust_tsp_2018_largelist,sahashemi_mcgill_tsp_2017_fastflexible,jlin_lehigh_tvlsi_2016_highthpt}
are also shown for comparison. When $\mathcal{L}_{l}$=32, the critical
path delay of $\mathbf{D}_{l}$ is larger than that of $\mathbf{D}_{s}$
so the clock frequency is determined by $\mathbf{D}_{l}$. Hence,
the clock frequency of $\mathbf{D}_{1}$ and $\mathbf{D}_{4}$ are
the same and lower than that of $\mathbf{D}_{5}$. The decoding throughput
of $\mathbf{D}_{4}$ is higher than that of $\mathbf{D}_{1}$ because
$\mathcal{L}_{s}$=1 is used and fewer clock cycles are required for
each frame in $\mathbf{D}_{4}$. When $\mathcal{L}_{l}$=8, the critical
datapath of $\mathbf{D}_{l}$ is shorter so the clock frequency of
$\mathbf{D}_{5}$ follows that of $\mathbf{D}_{s}$. The corresponding
throughput is the highest due to the high clock frequency. It is noted
that $M_{l}$=4, rather than $M_{l}$=8 which is used to maximize
the throughputs as reported in \cite{cxia_hkust_tsp_2018_largelist},
is used for all the $\mathbf{D}_{l}$. This is because the critical
path delay for $M_{l}$=8 is larger than those for $M_{l}$=4 and
also the one of $\mathbf{D}_{s}$. The clock frequency and hence the
throughput of $\mathbf{D}_{\text{TA}}$ will be lower if $M_{l}$=8
is used. 

From the area breakdown shown in \tabref \ref{tab:area_breakdown},
the area of the TA-SCL architecture is dominated by that of $\mathbf{D}_{l}$.
Numerically, $\mathbf{D}_{l}$ contributes 85\%, 88\% and 58\% of
the total area for $\mathbf{D}_{1}$, $\mathbf{D}_{4}$ and $\mathbf{D}_{5}$,
respectively. When $\mathcal{L}_{l}$ =32, the area overhead of $\mathbf{D}_{1}$
and $\mathbf{D}_{4}$ is 18\% and 11\% compared with the corresponding
single $\mathbf{D}_{l}$, respectively. The area of $\mathbf{D}_{4}$
with $\mathcal{L}_{s}$=1 is similar with that of $\mathbf{D}_{1}$
with $\mathcal{L}_{s}$=2 as the area is dominated by that of $\mathbf{D}_{l}$.
The area of $\mathbf{D}_{5}$ is much smaller than $\mathbf{D}_{1}$
and $\mathbf{D}_{4}$ because $\mathbf{D}_{l}$ with $\mathcal{L}_{l}$=8
has a much smaller area.

\tabref \ref{tab:syn_result} compares the synthesis results with
state-of-the-art architectures for polar codes with $N$=1024 and
$R\approx\frac{1}{2}$ \cite{cxia_hkust_tsp_2018_largelist,sahashemi_mcgill_tsp_2017_fastflexible,jlin_lehigh_tvlsi_2016_highthpt}.
Compared with the results in \cite{cxia_hkust_tsp_2018_largelist}
in which $M_{l}$=8 is used, the throughput gains achieved by $\mathbf{D}_{1}$
and $\mathbf{D}_{5}$ are 2.83x and 2.72x for $\mathcal{L}_{l}$=32
and 8, respectively, which are slightly lower than the theoretical
speed gains due to the clock rate issue as discussed above. For applications
just targeting at high SNR range, $\mathbf{D}_{4}$ can achieve a
throughput gain of up to 3.39x. Comparing $\mathbf{D}_{5}$ with the
decoders in \cite{sahashemi_mcgill_tsp_2017_fastflexible,jlin_lehigh_tvlsi_2016_highthpt}
with $\mathcal{L}_{l}$=8, the area is similar while the throughput
is nearly 4 times higher. The implementation results show that the
proposed TA-SCL architecture can significantly improve the decoding
throughput with a small hardware overhead and negligible error correction
performance degradation at a wide SNR range.

\section{Conclusion\label{sec:conclusion}}

In this work, a two-staged SCL decoding scheme is proposed, which
significantly increases the throughput of the polar decoding on hardware.
To analyse the error correction performance of TA-SCL decoding, a
mathematical model based on Markov chain is proposed. With a proper
selection of the design parameters, the performance loss is negligible
for a wide SNR range. A high-performance VLSI architecture is then
developed for the proposed TA-SCL decoding. Experimental results show
that the throughput of TA-SCL decoding implemented by the proposed
architecture are about three times as high as that of the state-of-the-art
architectures.

\appendices{}

\section{Proof of Proposition 2}

We first review Theorem 1 which is necessary to prove Proposition
2.

\noindent \textbf{Theorem 1. }\textit{(Bézout's identity) Let $a$
and $b$ be integers with greatest common divisor $d$. Then, there
exist integers $x$ and $y$ such that
\begin{equation}
ax+by=d.\label{eq:bezout}
\end{equation}
More generally, the integers of the form $ax$+$by$ are exactly the
multiples of $d$.}

Next, the proof of Proposition 2 is given as below.
\begin{IEEEproof}
We prove the irreduciblity of the Markov chain model first. First,
we consider the safe states only and try to prove that any safe state
$j$ is accessible from any other safe state $i$. Let $a$=$\beta_{n}$
and $b$=$\beta_{d}$. As $\beta_{n}\perp\beta_{d}$, $d$=1, and
\eqref{eq:bezout} can be rewritten as
\begin{eqnarray}
(\beta_{n}-\beta_{d})x+\beta_{d}(y+x) & = & 1\\
(\beta_{n}-\beta_{d})x'+\beta_{d}y' & = & 1\label{eq:bezout_tascl}
\end{eqnarray}
If $x'$ is positive and $y'$ is negative, \eqref{eq:bezout_tascl}
actually means state $i$+$\frac{1}{\beta_{d}}$ is accessible from
state $i$ after $(|x'|$+$|y'|)t_{s}$ during which $|x'|$ and $|y'|$
frames are decoded correctly and incorrectly by $\mathbf{D}_{s}$,
respectively. State $i$-$\frac{1}{\beta_{d}}$ is also accessible
as \eqref{eq:bezout_tascl} is still valid when its right-hand side
is -1 according to Theorem 1. Thus, any safe state $j$ can be accessed
from state $i$ by repeating this procedure. The accessibility of
a idle/hazard state from/to a safe state is obvious. Thus, it is possible
to get to any state from any state in this model and the irreduciblity
is proved. 

An irreducible Markov chain is aperiodic if any state is aperiodic.
As state 0 is aperiodic, the aperiodicity is proved.%
\end{IEEEproof}

\end{document}